\RequirePackage{ifpdf}
\documentclass[20pt, letterpaper, hyper]{JHEP3}

\pdfoutput=1

\usepackage[utf8]{inputenc}

\usepackage{epic,eepic}
\usepackage{amsmath,amssymb,amsfonts}
\usepackage{makeidx}

\usepackage{dsfont}
\usepackage{graphicx}
\usepackage{csquotes}
\usepackage{cite}

\usepackage{float}
\newlength \savewidth

\def\bea{\begin{eqnarray}}
\def\eea{\end{eqnarray}}

\def\i{\item} 
\def\mm{\mathcal}

\def\pa{\partial}

\def\a{\alpha}  \def\g{\gamma} \def\d{\delta}     \def\m{\mu}   \def\t{\theta} \def\k{\kappa} \def\l{\lambda}  
\def\G{\Gamma}  \def\D{\Delta}     \def\S{\Sigma} \def\O{\Omega}

\newcommand{\ben}{\begin{equation}}
\newcommand{\een}{\end{equation}}
\newcommand{\be}{\begin{equation}}
\newcommand{\ee}{\end{equation}}
\newcommand{\ba}{\begin{eqnarray}}
\newcommand{\ea}{\end{eqnarray}}

\newcommand{\beq}{\begin{equation}}
\newcommand{\eeq}{\end{equation}}
\newcommand{\beqa}{\begin{eqnarray}}
\newcommand{\eeqa}{\end{eqnarray}}
\newcommand{\beqar}{\begin{eqnarray*}}
	\newcommand{\eeqar}{\end{eqnarray*}}

\def\pa {\partial}

\def\t6 {T_\mt{D6}}

\newcommand{\mt}[1]{\textrm{\tiny #1}}

\def\cale         {{\cal E}}

\def\ee           {{\rm e}}

\def\sqr#1#2{{\vcenter{\vbox{\hrule height.#2pt
				\hbox{\vrule width.#2pt height#1pt \kern#1pt
					\vrule width.#2pt}\hrule height.#2pt}}}}

\def\a{\alpha}

\def\g{\gamma}
\def\ee{\cale}

\def\aa1{\phi}
\def\cc1{\psi}

\def\G{\Gamma}

\def\D{\Delta}
\def\k{\kappa}

\def\k{\kappa}

\title{\textbf{{\Large   On Volumes of Subregions in Holography and Complexity }}}
\author{Omer Ben-Ami${}^{1}$, Dean Carmi${}^{1,2}$
	\\
	${}^1$ \textit{Raymond and Beverly Sackler Faculty of Exact Sciences \\
		\ $\,$	School of Physics and Astronomy, Tel-Aviv University, Ramat-Aviv 69978, Israel}
\\
${}^2$ \textit{Perimeter Institute for Theoretical Physics\\
	\ $\,$	31 Caroline Street North, ON N2L 2Y5, Canada}
\\
	
}

\abstract{The volume of the region inside the bulk Ryu-Takayanagi surface is a codimension-one object, and a natural generalization of holographic complexity to the case of subregions in the boundary QFT. We focus on time-independent geometries, and study the properties of this volume in various circumstances. We derive a formula for computing the volume for a strip entangling surface and a general asymptotically AdS bulk geometry. For an AdS black hole geometry, the volume exhibits non-monotonic behaviour as a function of the size of the entangling region (unlike the behaviour of the entanglement entropy in this setup, which is monotonic). For setups in which the holographic entanglement entropy exhibits transitions in the bulk, such as global AdS black hole, geometries dual to confining theories and disjoint entangling surfaces, the corresponding volume exhibits a discontinuous finite jump at the transition point (and so do the volumes of the corresponding entanglement wedges). We compute this volume discontinuity in several examples. Lastly, we compute the codim-zero volume and the bulk action of the entanglement wedge for the case of a sphere entangling surface and pure AdS geometry.}	

\begin{document}

\tableofcontents	
\setcounter{footnote}{0}

\allowdisplaybreaks



\section{Introduction}	

Quantum physics differs from classical physics in two important aspects: entanglement and complexity. The entanglement entropy (EE) (e.g \cite{Calabrese:2004eu,Casini:2009sr,Bombelli:1986rw,Lewkowycz:2013nqa,Srednicki:1993im,Casini:2011kv,Holzhey:1994we,Solodukhin:2008dh,Ryu:2006ef,Ryu:2006bv,Nishioka:2009un,Headrick:2010zt,Hung:2011xb,Hung:2011ta,Ben-Ami:2015zsa,Blanco:2013joa,Nishioka:2006gr,Klebanov:2007ws}) is a measure of the quantum correlations of a quantum state, and is extremely useful in many quantum systems, ranging from condensed matter physics to black hole physics. The AdS/CFT correspondence \cite{Maldacena:1997re,Gubser:1998bc,Witten:1998qj,Aharony:1999ti} gives a simple geometrical way to compute entanglement entropy in holographic QFTs, using the Ryu-Takayanagi (RT) formula \cite {Ryu:2006ef,Ryu:2006bv} (and its covariant HRT generalization \cite{Hubeny:2007xt}):
\bea
S= \frac{\textrm{Area} (\g_A)}{4G_N}
\eea 
where $\g_A$ is the bulk extremal surface, and $G_N$ is Newton's constant.

However the entanglement (Renyi) entropies are not the whole story, and a quantum state contains a large amount of information which is not contained in them. Quantum complexity is a quantum information quantity which has to do with the difficulty converting one quantum state to another quantum state.
In the following we briefly review the concept of quantum complexity, and its proposed holographic conjectures as given in \cite{Susskind:2014moa,Brown:2015bva,Brown:2015lvg,Stanford:2014jda,Susskind:2014jwa,Susskind:2014rva,susskind,Brown:2016wib}.

Consider $K$ qubits in an arbitrary quantum state:
\bea
| \psi \rangle = \sum_{j_1, \ldots j_K}c_{j_1, \ldots j_K} | j_1, \ldots j_K \rangle
\eea 
where the $c_{j_1, \ldots j_K}$ are complex numbers, and each $j_i$ is either 0 or 1. It takes a huge amount of information, namely $2^K$ complex numbers, to specify the state of $K$ qubits. To define quantum complexity, one starts by considering some simple reference state $ | 0\ldots 0\rangle$, which is a product state. One then chooses a universal gate set of 2-qubit unitary operations. Every state in the Hilbert can be approached (arbitrarily close) by applying a sequence of these 2-qubit operations. The quantum complexity of a state $| \psi \rangle$ is then defined as the minimal number of 2-qubit gates required to prepare the state $| \psi \rangle$ from the reference state $| 0\ldots 0\rangle$. When $K$ is large, nearly all states $|\psi \rangle$ have a maximal complexity which is exponential in $K$:
\bea
\mm{C}_{max} \sim e^K
\eea 

On the other hand, the maximal entropy is $K \log 2$ and thus the maximal complexity is exponentially larger then the maximal entropy. For a system which undergoes thermalization, the quantum complexity grows linearly in $t$ for a very long time which is much larger than the thermalization time. After an exponentially large time $t \sim e^K$, the complexity saturates at a constant value $\mm{C}_{max}$. There can be fluctuations around the mean value $\mm{C}_{max}$, and the time at which the system can fluctuate back to the reference state is doubly exponential: $t_{rec}\sim e^{e^K}$. It was also argued that the rate of change of complexity is proportional to the entropy and the temperature of the system: $\frac{d\mm{C}}{dt} =TS$.

\begin{figure}[!h]
	\centering
	
	\begin{minipage}{0.48\textwidth}
		\centering
		
		\includegraphics[scale=0.3]{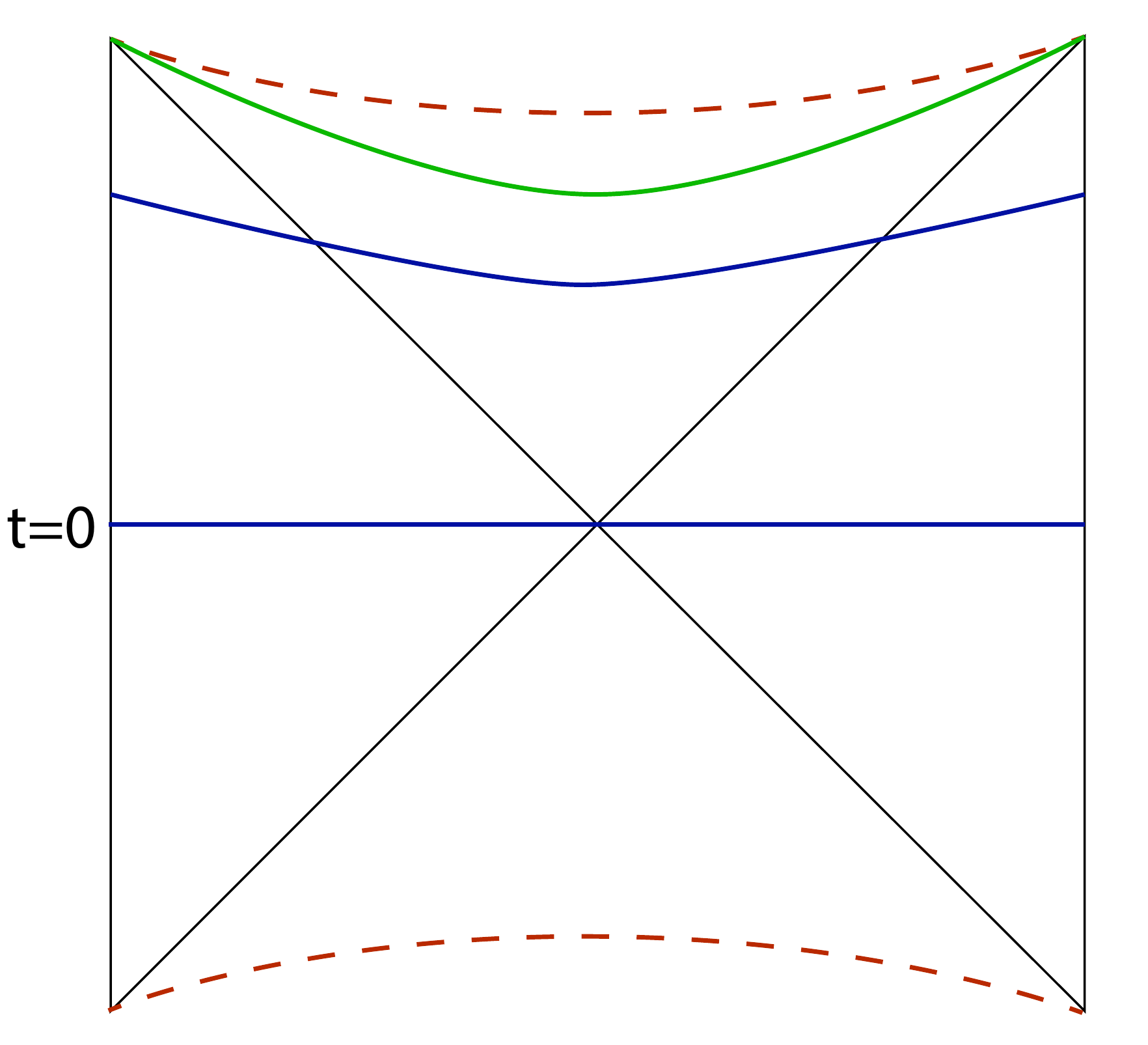}
	\end{minipage}
	\begin{minipage}{0.48\textwidth}
		\centering 
		\includegraphics[scale=0.25]{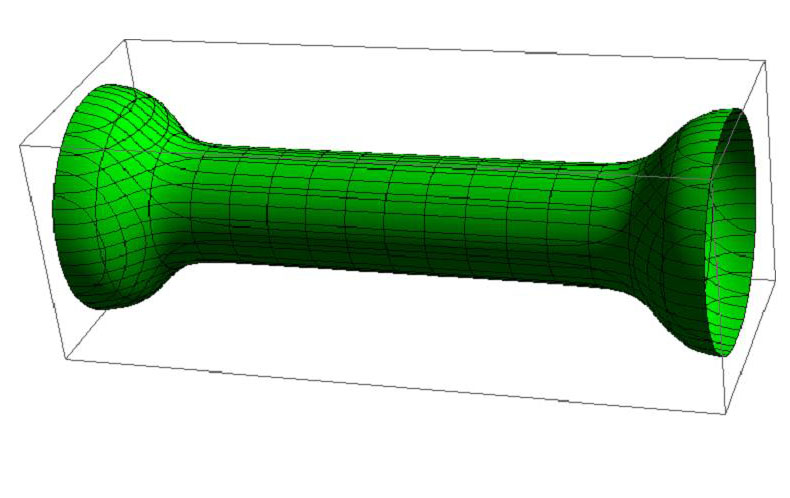}
	\end{minipage}
	\caption{\textbf{Left:} Illustration of the double sided AdS black hole Penrose diagram dual to the thermofield double state. The left and right boundaries are where $CFT_L$ and $CFT_R$ live. The singularity is shown in dashed red. A few maximal codim-one surfaces are shown, including the one at $t\to \infty$ in green (which does not reach the singularity.). \textbf{Right:} Illustration of embedding diagram of a wormhole.}
		\label{fig:space}
\end{figure}

Consider now the double sided AdS black hole geometry \cite{Maldacena:2001kr} which is dual to a specific entangled state, namely the thermo-field double state, see Fig~\ref{fig:space}. The left and right CFTs can be viewed as being connected  through the black hole interior by a wormhole (e.g by drawing a space-like slice between the two sides)\footnote{The relation between the entanglement (EPR) of causally disconnected systems and the wormhole (Einstein-Rosen bridge) in the dual bulk theory was conjectured by \cite{Maldacena:2013xja} to hold more generally and is now known as ER=EPR.}. The wormhole is a dynamical object which grows linearly in time at late times. In this classical bulk geometry the wormhole continues to grow forever. On the other hand, thermal equilibrium is achieved much quicker, and the entanglement entropies saturate \cite{Hartman:2013qma,Hubeny:2013dea,Liu:2013iza,Liu:2013qca}.

In \cite{Susskind:2014moa,Brown:2015bva,Brown:2015lvg,Stanford:2014jda,Susskind:2014jwa,Susskind:2014rva,susskind,Brown:2016wib} there were two important conjectures made about how to compute complexity in holography\footnote{For additional recent work on complexity, see \cite{Cai:2016xho,Barbon:2015ria,Barbon:2015soa,Chemissany:2016qqq,Vanchurin:2016met,Lehner:2016vdi,divergences}.}. The proposals relate quantum complexity to the size of the wormhole. The linear growth of complexity in a thermalizing system is then holographically dual to the linear growth of the wormhole in an AdS black hole geometry. More precisely, the first conjecture states that the complexity is given by the volume\footnote{The volumes of the interior of black holes in flat space were studied in \cite{Christodoulou:2014yia,Christodoulou:2016tuu}.} of the codim-one maximal bulk surface that ends on the boundary at a time $t$:
\bea
\mm{C}= \frac{V}{G \ell}
\eea 
where $\ell$ is some length scale (typically the AdS radius or the BH radius) which needs to be chosen for each case at hand. This is somewhat unsatisfactory, and led  \cite{Brown:2015bva,Brown:2015lvg} to suggest a refined version where the dual of complexity equals the action evaluated in a specific bulk region, as we review below. 

For AdS black holes the collection of all maximal slices foliate the entire space-time outside of the horizon, but not the entire interior of the black hole. At $t\to \infty$ the final slice is achieved, and it does not reach the singularity.
 However, the volume grows linearly in $t$ (for late times) as expected from quantum complexity. Additionally, the volume passes a few tests such as having the correct behavior in shock wave geometries.

The second conjecture \cite{Brown:2015bva,Brown:2015lvg} states that the complexity is given by the bulk action evaluated on the Wheeler-deWitt patch attached at some boundary time $t$\footnote{The WDW patch is the union of all spatial curves anchored at a time $t$ on the boundary.}:
\bea
\mm{C}= \frac{\mm{A}}{\pi \hbar}
\eea
	
The action proposal is more satisfactory than the volume proposal in the sense that one does not need to choose by hand a length scale $\ell$.
The calculation of the action on the WDW patch has a few challenging features: surface terms arising from the null boundaries of the WDW patch, corner terms where the null boundaries meet, and a surface term from the singularity of the black hole\footnote{Some of these issues can be ignored when calculating the time derivative of the WDW action.}, see \cite{Brown:2015bva,Brown:2015lvg,Lehner:2016vdi,divergences}. 

In \cite{Brown:2015bva,Brown:2015lvg} it was shown that for AdS black holes the rate of change of the WDW action at late times is:
\bea
\frac{d \mm{A}}{d t} = 2M
\eea 

This result led to the very interesting conjecture \cite{Brown:2015bva,Brown:2015lvg,lloyd} that black holes saturate the inequality:
\bea
\frac{d\mm{C}}{dt}\leq \frac{2E}{\pi \hbar}
\eea 
and as a result, black holes are the fastest computers in nature.

The recent progress in holographic complexity, and especially the complexity equals volume conjecture, serves us as motivation to study bulk codim-one surfaces and their volumes. Volumes of codim-one surfaces can be defined also for subregions on the boundary, and can be seen as a natural subregion generalization of holographic complexity. In \cite{Alishahiha:2015rta} the codim-one volume contained inside the codim-two Ryu-Takyanagi surface was considered as the dual of complexity for subregions\footnote{For additional related work , see \cite{MIyaji:2015mia,Lashkari:2015hha,Momeni:2016ekm,Momeni:2016qfv,Momeni:2016yts,Mazhari:2016yng}}. In our work we further study codim-one volumes for subregions, and uncover some interesting properties which they possess. We will also be agnostic about the length scale $\ell$ and assume that one can be chosen appropriately. The volume is also closely related to the volume of the entanglement wedge of the subregion, which is believed to be the region in the bulk corresponding to the reduced density matrix of the subregion \cite{Headrick:2014cta}. The entanglement wedge is the domain of dependence of the ``subregion volume". 
	
The contents of this paper are as follows. In section~\ref{sec:divergence} we consider the codim-one volume inside the Ryu-Takayanagi surface as a natural dual to ``sub-region complexity". We proceed to compute this volume for time-independent geometries, and we write a general formula for the case of a strip entangling surface. We compute the volume in a few examples, and plot the temperature dependence of the volume for an AdS black hole geometry which exhibits non-monotonic behavior. In section~\ref{sec:voldisc} we show that the volume has a discontinuous jump in geometries in which there is a transition between 2 bulk minimal surfaces. We compute this jump in several examples. In section~\ref{sec:discuss} we summarize our results, and discuss future directions. In Appendix~\ref{sec:sphere} we present the computation of the leading temperature correction for an AdS black hole geometry and sphere entangling surface. In Appendix~\ref{sec:wedge3} we calculate the volume and action of the entanglement wedge for the case of a sphere entangling surface and pure AdS.


\section{Bulk Subregion Volumes}
\label{sec:divergence}

Consider a holographic QFT in $d$ space-time dimensions and an entangling surface of typical length $R$. Now consider the codim-one volume contained inside the corresponding Ryu-Takayanagi surface. The divergence structure of this volume was shown to be\footnote{We set $L_{\textrm{AdS}}=G_N=\ell=1$ such that $V$ is dimensionless.} (see \cite{divergences,Alishahiha:2015rta}):
\bea
\label{eq:lknnd}
V = c_{d-1}\frac{R^{d-1}}{\d^{d-1}} + c_{d-3}\frac{R^{d-3}}{\d^{d-3}} + \ldots + 
\begin{Bmatrix}
	c_1\frac{R}{\d} +(-1)^{\frac{d-1}{2}}c_0\ \ \ \ \ \ \ , \ \ \ \ d=even \\ 	c_2 \frac{R^2}{\d^2} +(-1)^{\frac{d-2}{2}}\tilde{c}_0\log(\frac{R}{\d})  \ \ \ , \ \ \ d=odd
\end{Bmatrix}
\eea 
where $\d$ is the UV cutoff. The leading divergence scales like the volume of the entangling region, and there are subleading divergences. The expansion above contains universal terms, meaning in odd $d$ there is a universal logarithmic divergence, and in even $d$ there is universal finite term (i.e opposite to the EE case). 

In \cite{divergences} we define a covariant generalization of the volume inside a Ryu-Takayanagi surface, which is applicable also to time-dependent geometries. The prescription is the following: 
\begin{itemize}
	\i For a given time $t$ on the boundary and an entangling region $A$, find the codim-two HRT surface \cite{Hubeny:2007xt}. Then find the codim-one maximal volume attached to the HRT surface and the entangling region $A$. This gives a covariant construction of the volume associated with a subregion on the boundary. 
\end{itemize}

 Similar to the case of entanglement entropy \cite{Hung:2011ta,Hung:2011xb}, in \cite{divergences} this divergence structure is written in terms of the curvatures of the boundary CFT.

\subsection{Strip Entangling surface}	
\label{sec:strip}

\begin{figure}[!h]
	\centering
	\includegraphics[width=70mm]{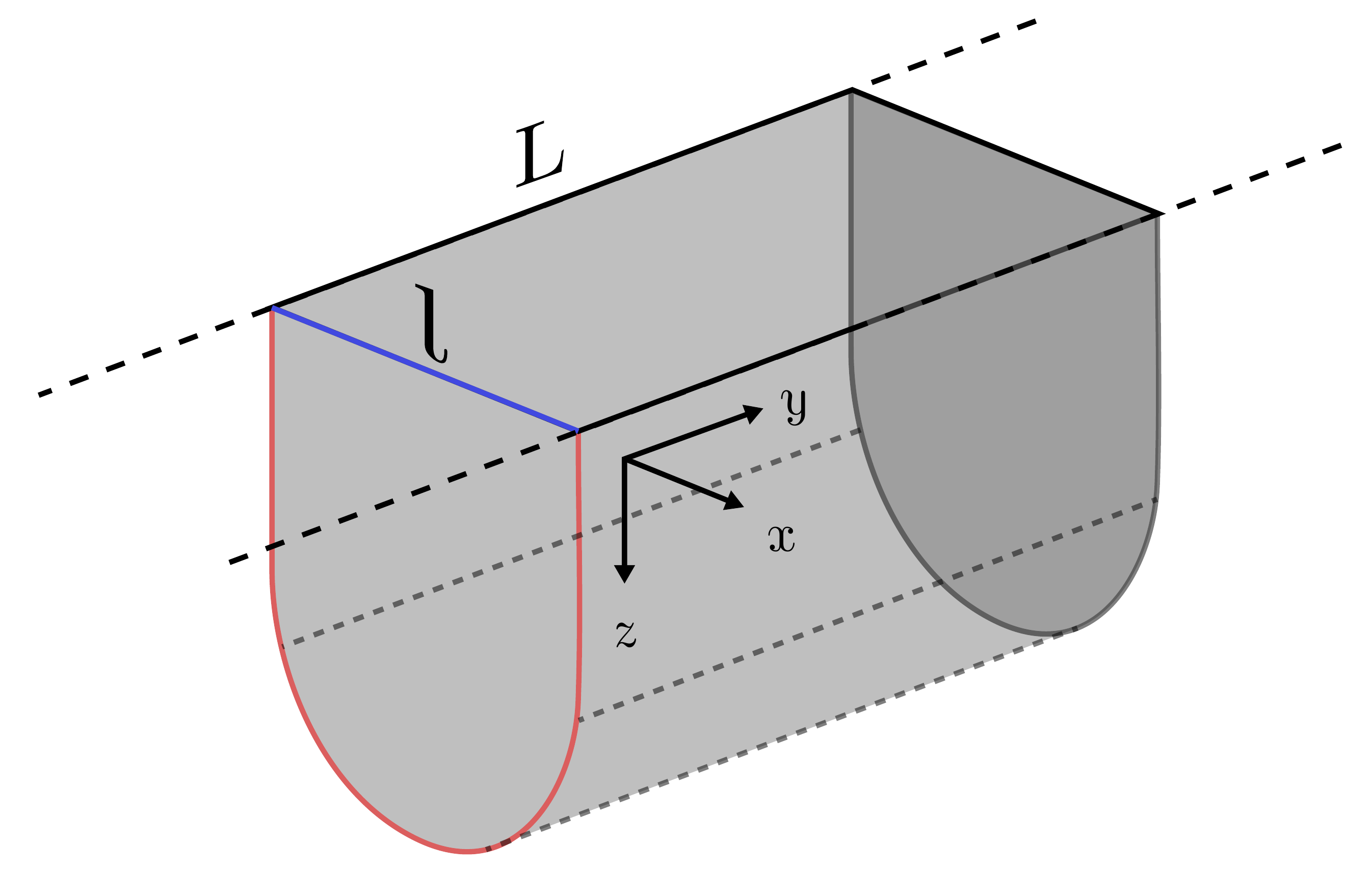}	
	\caption{A strip entangling region with width $l$ and length $L\to \infty$ on the boundary, and its corresponding minimal surface in the bulk. The boundary is at $z\rightarrow 0$.\label{strip}}
\end{figure}

In this section we study the volume prescription written above for the case of a strip entangling surface of length $l$ and width $L \to \infty$, in a time-independent background geometry\footnote{The translational symmetry of the strip considerably simplifies this problem.}, see Fig.~\ref{strip}. Namely we compute the volume inside the Ryu-Takyanagi surface. The case of a sphere in pure AdS was studied in \cite{Alishahiha:2015rta}. Since the strip entangling surface has zero curvature, there will only be a single divergence (``volume law") plus a finite term:
\bea
V = c_{d-1}\frac{lL^{d-2}}{\d^{d-1}} + c_{finite}
\eea

Now consider a general asymptotically $\textrm{AdS}_{d+1}$ metric:
\bea
ds^2 =\frac{1}{z^2}\Big[ f_0(z)dt^2 + f_1(z) dx_\m^2 + f_2(z) dz^2  \Big] 
\eea
where $f_0(z)$, $f_1(z)$, and $f_2(z)$ are some arbitrary functions of $z$, which at the boundary are $f_i(z=0)=1$.

The holographic entanglement entropy (area of the minimal surface) of a strip of length $l$ is:
\bea
\label{eq:jmsoskk1}
S(z_*) = \frac{2L^{d-2}}{4G_N}   \int_{0}^{l/2} dx\frac{f_1^{\frac{d-2}{2}}}{z^{d-1}}\sqrt{f_1(z)+f_2(z)(\pa_x z)^2}=
\frac{2L^{d-2}}{4G_N}   \int_{\d}^{z_*} \frac{dz}{z^{d-1}} \frac{\sqrt{f_2 f_1^{d-2}}}{\sqrt{1- \frac{f^{d-1}_1(z_*)z^{2d-2}}{f^{d-1}_1(z)z_*^{2d-2}}}}
\eea

where $z_*$ is the deepest point of the minimal surface in the bulk. In the second equation the corresponding equation of motion was used. The equation of motion is:

\bea
\label{eq:eomfhh}
\pa_x z = \mp \sqrt{\frac{f_1}{f_2}}\sqrt{\frac{f^{d-1}_1(z)z_*^{2d-2}}{f_1^{d-1}(z_*) z^{2d-2}}-1} \ \ \ \ \ \ \ \ \Rightarrow \ \ \ \ \ \ \ \ \ \ x_1(z)= \int_{z}^{z_*} dZ \frac{\sqrt{\frac{f_2(Z)}{f_1(Z)}}}{\sqrt{\frac{f^{d-1}_1(Z)z_*^{2d-2}}{f_1^{d-1}(z_*) Z^{2d-2}}-1}}
\eea 
Note that the solution for $x_1(z)$ is computed as a one-dimensional integral.

Since the background is static, the volume inside the minimal surface is obtained simply by integrating the inside of the minimal surface. We do this integral by slicing the bulk with planes of constant $z$, with the result:
\bea
\label{eq:1}
V(z_*)=    \int dx \int d^{d-2}y \sqrt{\tilde g} =  2L^{d-2}   \int_{\d}^{z_*} \frac{dz \sqrt{f_1^{d-1} f_2}}{z^{d}} \int_0^{x_1(z)} dx
=  2L^{d-2}  \int_{\d}^{z_*} \frac{dz \sqrt{f_1^{d-1}f_2}}{z^{d}}  x_1(z)
\eea
where $\sqrt{\tilde g}$ is the volume element, and $x_1(z)$ is the profile of the minimal surface given in Eq.~\ref{eq:eomfhh}.
Plugging Eq.~\ref{eq:eomfhh} into Eq.~\ref{eq:1} gives:
\bea
\label{eq:jmsoskk2}
	V(z_*)= 2L^{d-2} \int_{\d}^{z_*} \frac{dz \sqrt{f_1^{d-1}(z) f_2(z)}}{z^{d}} \int_{z}^{z_*} dZ\frac{\sqrt{\frac{f_2(Z)}{f_1(Z)}}}{\sqrt{\frac{f^{d-1}_1(Z)Z_*^{2d-2}}{f_1^{d-1}(Z_*) Z^{2d-2}}-1}}
\eea
This is our formula for the volume as a function of $z_*$ for any asymptotically AdS geometry (any $f_{1,2,3}(z)$).
This formula is only slightly more complicated than the corresponding formula for the entanglement entropy of the strip Eq.~\ref{eq:jmsoskk1}, since there are two integrals that one needs to perform instead of one. 

From Eq.~\ref{eq:eomfhh} we have:
\bea
\label{eq:jmsoskk3}
l(z_*) = 2\int_{\d}^{z_*} dz \frac{\sqrt{\frac{f_2(z)}{f_1(z)}}}{\sqrt{\frac{f^{d-1}_1(z)z_*^{2d-2}}{f_1^{d-1}(z_*) z^{2d-2}}-1}}
\eea
So in Eqs.~\ref{eq:jmsoskk2} and \ref{eq:jmsoskk3} we have formulas for $V(z_*)$ and $l(z_*)$, and we can plot $V$ as a function of $l$ for any background.
In the following we will compute the volume in several examples using these formulas.

Note that there is another solution to the equation of motion for which $x_1(z)={l\over 2}$, meaning a constant solution for which the surface does not cap off smoothly at some point $z_*$ in the bulk. Although this is not the minimal surface for most cases\footnote{As we will review below, for confining theories the constant solution can be the minimal surface.}, it will be interesting to consider its behavior compared to the minimal surface.  The volume of this surface is given by:
\bea
\label{eq:jd9f4}
V_c(l)=  lL^{d-2}   \int_{\d}^{\infty} \frac{dz \sqrt{f_1^{d-1} f_2}}{z^{d}} 
\eea
\subsubsection{Pure AdS}

For a pure $\textrm{AdS}_{d+1}$ background $f_i(z)=1$, the EE is given by:
\begin{equation}
\label{eq:EEonestripAdS}
S(l)={1\over 2(d-2)}\left(L\over \delta \right)^{d-2}-{2^{d-3}\pi^{d-1 \over 2}\over d-2}\left(\Gamma\left(d\over 2d-2\right)\over \Gamma\left(1\over d-2\right)\right)^{d-1}\left(L\over l\right)^{d-2}
\end{equation}

Similarly, the volume is:
\bea
\label{eq:kjbbbb}
V(l)= \frac{1}{d-1}\frac{L^{d-2}l}{\d^{d-1}} - \frac{2^{d-2}\pi^{\frac{d-1}{2}}\G^{d-3}(\frac{d}{2d-2})}{(d-1)^2\G^{d-3}(\frac{1}{2d-2})} \frac{L^{d-2}}{l^{d-2}} \equiv \frac{c_1}{\d^{d-1}} - \frac{c_0}{l^{d-2}}
\eea 
Note that the finite term has the same $l$ dependence as the finite part of the entanglement entropy. 

Similarly, the volume of the constant solution Eq, ~\ref{eq:jd9f4} is given by:
\begin{equation}
V_c(l)=\frac{1}{d-1}\frac{L^{d-2}l}{\d^{d-1}}
\end{equation}
we get that for this solution there is no finite, universal, term.

\subsubsection{$D_p$-branes}

Consider a $D_p$-brane background with metric \cite{Itzhaki:1998dd}:
\begin{align}
ds^2&=\alpha'\left({U^{(7-p)\over 2}\over c_p}\left(-dt^2+dx_1^2 +\cdot+dx_{p}^2\right)+{c_p\over U^{(7-p)\over 2}}dU^2+c_pU^{(p-3)\over 2}d\Omega^2_{8-p}\right)\\\nonumber
e^\phi&=(2\pi)^{2-p}g_{\textrm{YM}}^2\left({c_p^2\over U^{7-p}}\right)^{3-p \over 4}
\end{align}
where $c_p=g_{\textrm{YM}}N^{1\over 2}\sqrt{2^{7-2p}\pi^{9-3p\over 2}\Gamma\left({7-p\over 2}\right)}$

For non-conformal theories, in which the dilaton and the volume of the internal coordinates are in general not constant, we use the natural generalization from quantum gravity for the codim-2 surface:
\begin{equation}
S_A={1\over 4G_N^{(10)}}\int d^8\sigma e^{-2\phi}\sqrt{G^{(8)}_\textrm{ind}}
\end{equation}
Similarly, since we expect the general relation between complexity and the codim-1 surface $V\sim {1\over G_N^{(d+2)}\ell}$, the generalization will be:
\begin{equation}
V\sim{1\over \ell G_N^{(10)}}\int d^9\sigma e^{-2\phi}\sqrt{G^{(9)}_\textrm{ind}}
\end{equation}

Plugging in the equation for the dilaton and the metric we get that for the minimal surface\footnote{To avoid clutter we set $c_p=1$. }:
\begin{equation}
{dU\over dx}=U^{7-p \over 2}\sqrt{{U^{9-p}\over U^{9-p}_*}-1}
\end{equation}
where $U_*$ is the turning point of the surface, and we assume that we can trust the supergravity solution at least up to this region.

Following the same procedure as for the conformal case we can extract the dependence of the finite term on the length of the strip.
The corresponding volume inside the minimal surface is:
\bea
\label{eq:pkkkde1}
V_{\textrm{finite}}\sim  -{f\left(p,g_{\textrm{YM}},N\right) \over l^{p+5\over 2(5-p)}}
\eea 
where $f\left(p,g_{\textrm{YM}},N\right)$ is some function which depends on the dimensions of the D$p$ brane, the coupling constant and the central charge.
Like for AdS, the finite term is an inverse power law $- l^\frac{-(p+5)}{2(5-p)}$. However for the $D_p$ brane case the power in the volume finite term is different from the power in the entanglement entropy, which has the form $S_{finite}\sim - l^{-\frac{4}{5-p}}$.

\subsubsection{The leading temperature correction to the volume}	

In this section we calculate the 1st order temperature correction to the volume in an AdS black hole geometry. Consider again the metric:
\bea
ds^2 =\frac{1}{z^2}\Big[ f_0(z)dt^2 + f_1(z) dx_\m^2 + f_2(z) dz^2  \Big] 
\eea
The form of the temperature perturbation is $f_2(z)=\frac{1}{f_0(z)} = \frac{1}{1-k_0T^d z^{d}} \approx 1+k_0T^dz^{d}$, where $k_0$ is a constant, and $T$ is the temperature. 
The finite term of the volume will have an expansion of the form\footnote{For temperature expansions for the holographic entanglement entropy see \cite{Fischler:2012ca}.}:
\bea
V_{finite}= \Big(\frac{L}{l}\Big)^{d-2}\Big[ a_1+a_2(Tl)^d + a_3(Tl)^{2d}+ \dots  \Big]
\eea 
and we want to find the first correction $a_2$.
 We use Eqs.~\ref{eq:jmsoskk2} and \ref{eq:jmsoskk3}, and after a few manipulations we get an integral expression for the volume:
\begin{align}
\label{eq:volks}
\D V \equiv V-V_{CFT} &=
c_0T^dl^{2}(2g_0)^{-3} \bigg[-g_1 g_0 + \bigg.\\\nonumber
&\bigg. 
\int_{1}^{\infty} dU U^{d-1} \left(   - dg_1 g_0    
+ (d-1)g_1   g_0( U)   +
g_0  U^{-d-1}   g_0( U) +
g_0    g_1( U) \right) \bigg]
\end{align}

where we subtracted the volume in the vacuum state: $V_{CFT}=V(T=0)$. 
We also defined the integrals:
\bea
g_0( U) \equiv \int_{1}^{ U}  \frac{dt}{t^2 \sqrt{t^{2d}-1}} \ \ \ \ , \ \ \ \ g_1( U) \equiv \int_{1}^{ U}  \frac{dt t^{-d-3}}{\sqrt{t^{2d}-1}}
\eea 

and

\bea
g_0 \equiv g_0( U\to \infty )  \ \ \ , \ \ \ \ g_1 \equiv g_1( U\to \infty )
\eea

\begin{figure}[!h]
	\centering
	
	\begin{minipage}{0.48\textwidth}
		\centering
		
		\includegraphics[width=63mm]{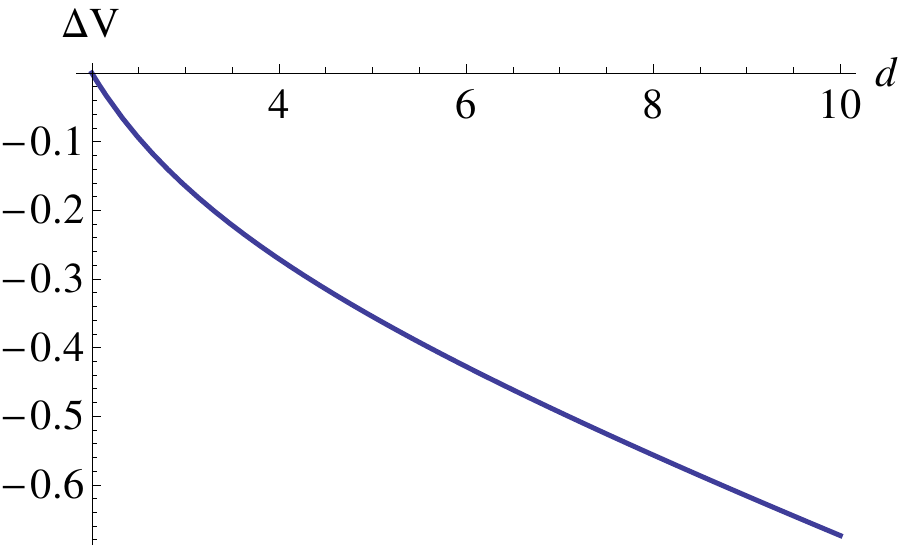}
	\end{minipage}
	\begin{minipage}{0.48\textwidth}
		\centering 
		\includegraphics[width= 63mm]{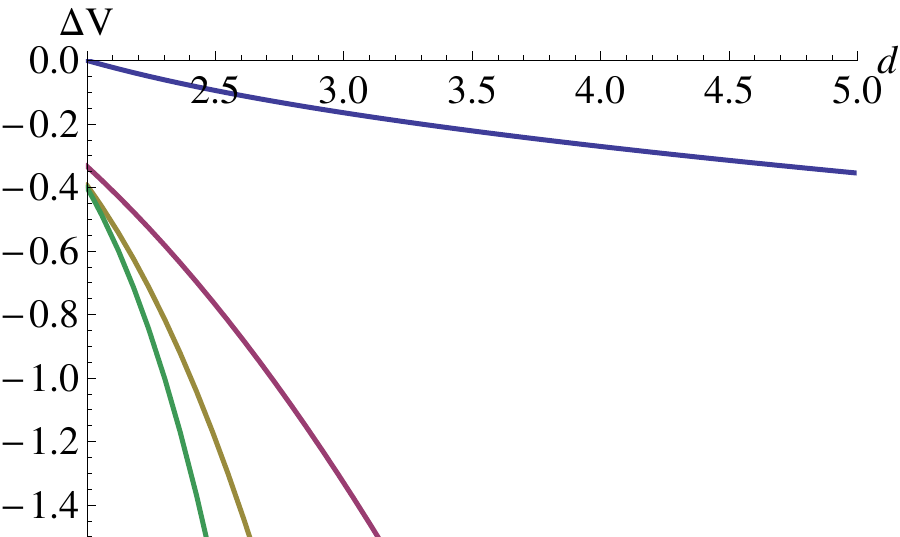}
	\end{minipage}
	\caption{\textbf{Left:} Plot of $\D V \equiv V-V_{CFT}$ as a function of $d$ for a temperature perturbation. \textbf{Right:} Plot of $V-V_{CFT}$ as a function of $d$ for perturbations of the form $f_2(z)= 1+m z^{q}$. Curves from top to bottom correspond to: $q=d$, $q=d+1$, $q=d+2$, and $q=d+3$. \label{3}}
\end{figure}

We plot the leading correction $\D V$ from Eq.~\ref{eq:volks} as a function of the dimension $d$ in Fig.~\ref{3}-Left.
We see from the plot that the 1st order correction is negative for any $d$. This is different from the case of sphere (see appendix~\ref{sec:sphere} and \cite{Alishahiha:2015rta}), where the 1st order correction vanishes but the 2nd order is positive \cite{Alishahiha:2015rta}. For $d=2$ the correction vanishes as expected since the interval is like a one dimensional sphere.

Now consider a more general perturbation of the form $f_2(z)= 1+M z^{q}$. We again go through a similar calculation, and we show the results in
Fig.~\ref{3}-Right for several values of $q$. We see that the first order $\D V$ is negative for all $d$ and $q$. We also see that for $d=2$ the leading correction is non zero when $q\neq 2$. This implies that the vanishing leading correction for the sphere (obtained in \cite{Alishahiha:2015rta} and shown in section~\ref{sec:sphere}) applies only to temperature perturbations, and for a general perturbation the 1st order correction will not vanish.

%
%


	
%

\subsubsection{The volume at finite temperature}	

\begin{figure}[!h]
	\centering
	\begin{minipage}{0.48\textwidth}
		\centering		
		\includegraphics[width=75mm]{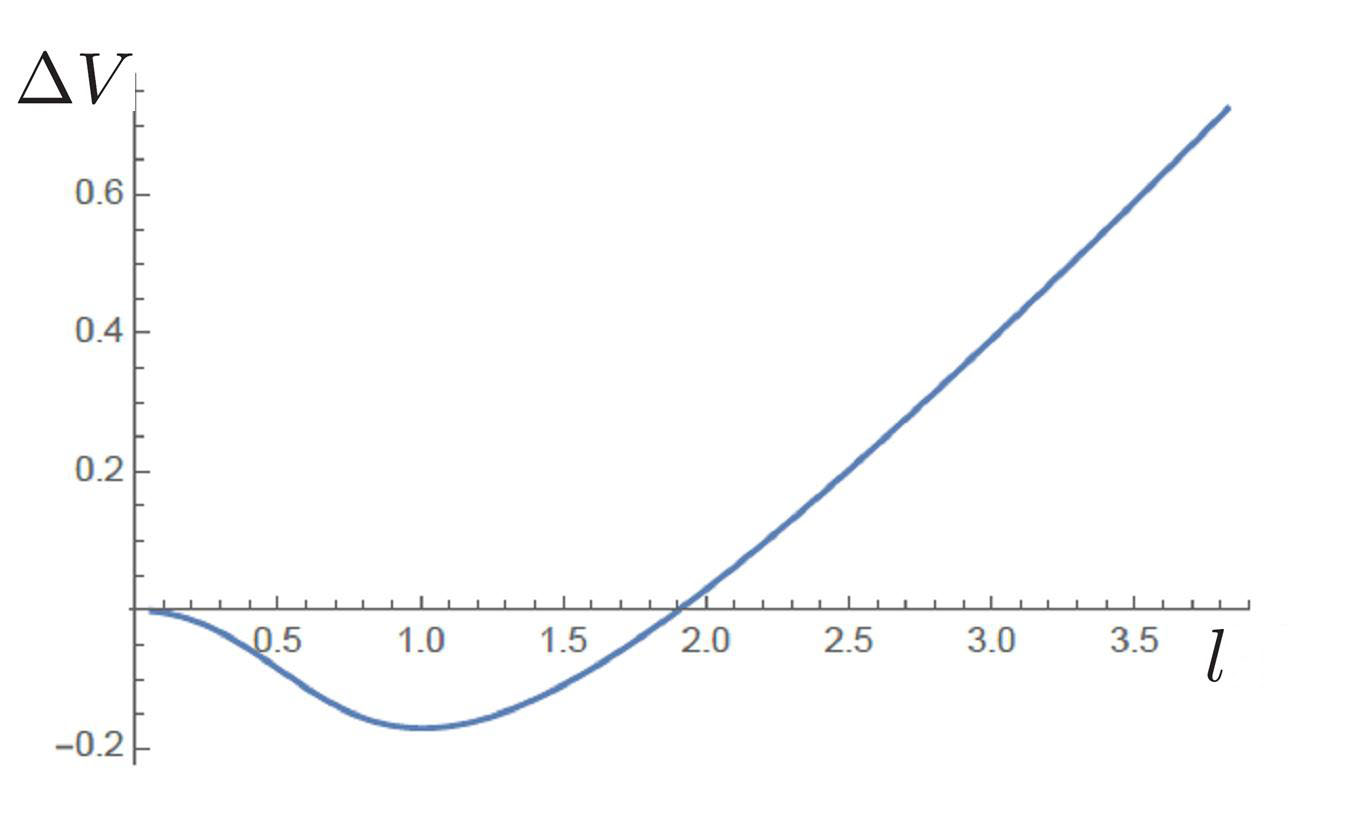}	
	\end{minipage}
	\begin{minipage}{0.48\textwidth}
		\centering
		\includegraphics[width= 80mm]{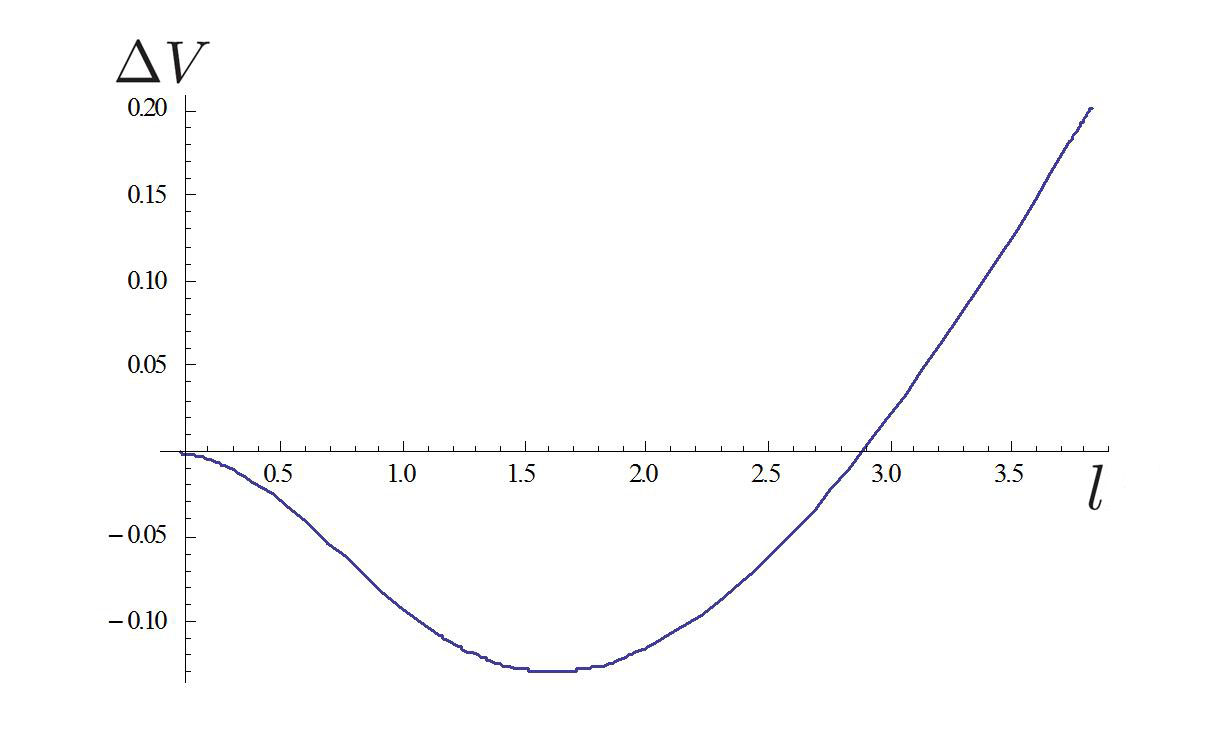}
	\end{minipage}
	\caption{Plots of $\D V \equiv V-V_{CFT}$ as a function of $l$ (we change $l$ and leave $T$ fixed). At $l=0$ we have $\D V=0$. \textbf{Left:} The $AdS_5$ black hole. \textbf{Right:} The $AdS_4$ black hole. \label{T1}}
\end{figure}

In the previous section we computed the 1st order temperature correction to the volume in various dimensions. In the high temperature regime we expect an expansion of the form\footnote{For a general shaped entangling surface the high temperature expansion will go like $V \sim a_0vT^{d-1}$, where $v$ is the volume of the entangling region. This is similar to the EE at high temperature.}:
\bea
V_{finite}= a_0L^{d-2}lT^{d-1}[1+O(1/(Tl))]
\eea 
which is linear in $l$ at very high $T$.
We now focus on the $AdS_4$ and $AdS_5$ planar black hole geometries and a strip entangling surface. We compute numerically the dependence of the volume on the strip's length at a finite constant temperature. Fig.~\ref{T1} shows the result of $\D V \equiv V-V_{CFT}$ as a function of the dimensionless parameter $l$ when $T$ is held fixed.

\begin{figure}[!h]
	\centering
		\begin{minipage}{0.48\textwidth}
		\centering		
		\includegraphics[width=80mm]{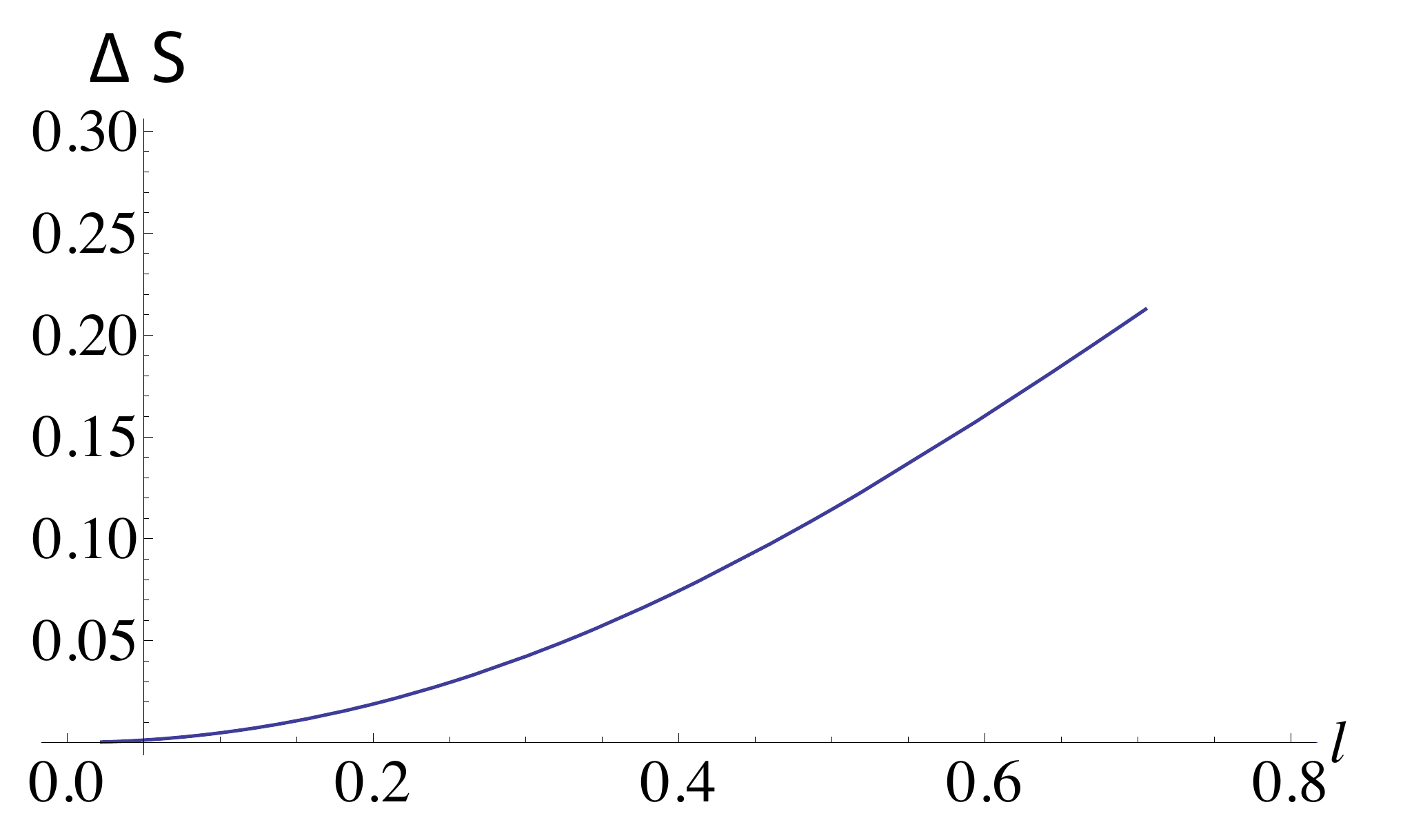}	
	\end{minipage}
	\caption{
		The entanglement entropy entropy $\D S = S -S_{CFT}$ as a function of $l$ (we change $l$ and leave $T$ fixed) for an $AdS_5$ black hole and a strip entangling surface. $\D S$  starts at the origin, and has linear behavior at large $lT$. \label{EE1}}
\end{figure}

The plot starts at the origin at $l=0$ and $\D V=0$, and we see the expected linear growth behavior at large $l$. Curiously the plot is non-monotonic and has a minimum. Compare this plot to the monotonic behavior of the entanglement entropy, shown in Fig.~\ref{EE1} for the $AdS_5$ black hole and a strip entangling surface.
It would be interesting to understand the meaning of the minimum in the volume. It would also be interesting to understand if for general shaped entangling surfaces the volume likewise has a minimum.

It is also easy to see that for the constant solution Eq. ~\ref{eq:jd9f4}, we have:
\begin{equation}
\Delta V_c\sim a_0L^{d-2}lT^{d-1}
\end{equation}
which \itshape{is} \normalfont monotonic and linear in $l$.



\subsubsection{$m$-strips in an AdS black hole geometry with large temperature}	

For an AdS black hole geometry at very large temperature and a strip entangling surface, the volume is linear in $l$, as we saw in the previous subsection. Now consider $m$ strips in an AdS planar black hole background, e.g \cite{Ben-Ami:2014gsa}. The length of the strips is denoted by $l$, and the distance between them $x$. Now take the separation between the strips to be very large: $xT >>1$. The bulk minimal surface will correspond to $m$ disconnected bulk surfaces. Nevertheless, in this section we compute the volume inside the \underline{connected} surface. Although it is not the minimal surface it has an interesting property. The volume of this surface is easily computed:
\bea
V_{connected}= C_0 (ml+(m-1)x) -C_0(m-1)x=  C_0 ml
\eea  	

where $C_0$ is a constant. We see that if we take the length of the strips $l$ to be constant and we separate the strips a large distance $x$ apart, then the volume inside the connected bulk minimal surface goes to a constant $C_0 ml$. If we take the ``mutual volume" by subtracting $m V_{1-strip}$, we simply get zero.

Compare this with the area (entanglement entropy) of such a connected surface:
\bea
S_{connected} \sim 2C_1  ((m-1)x +ml)  \quad   , \quad xT>>1
\eea  
and this quantity grows linearly with $x$.
So the volume of the connected surface goes to a constant independent of $x$, even though its area depends linearly on $x$.


\section{Volume Transitions and Discontinuities}
\label{sec:voldisc}

There are various cases in which the holographic entanglement entropy exhibits ``phase transitions", as one changes some parameter of the system. This occurs when two bulk minimal surfaces exchange dominance, and is reminiscent of 1st order phase transitions. The entanglement entropy is continuous across the transition, but there is a jump in the derivative of EE. Such transitions occur for example in disjoint entangling surfaces, global AdS black hole geometries, and confining geometries.

In this section we compute the behavior of the volume inside the minimal surface, when there are bulk transitions. In all of the examples that we consider, the volume has a discontinuous jump at the transition point. Note that these transitions imply that the volume of the entanglement wedge jumps discontinuously. In the following we will compute a few examples where such transitions occur.


\subsection{Two strips in $\textrm{AdS}_{d+1}$}

\begin{figure}[!h]
	\centering
	\includegraphics[width=120mm]{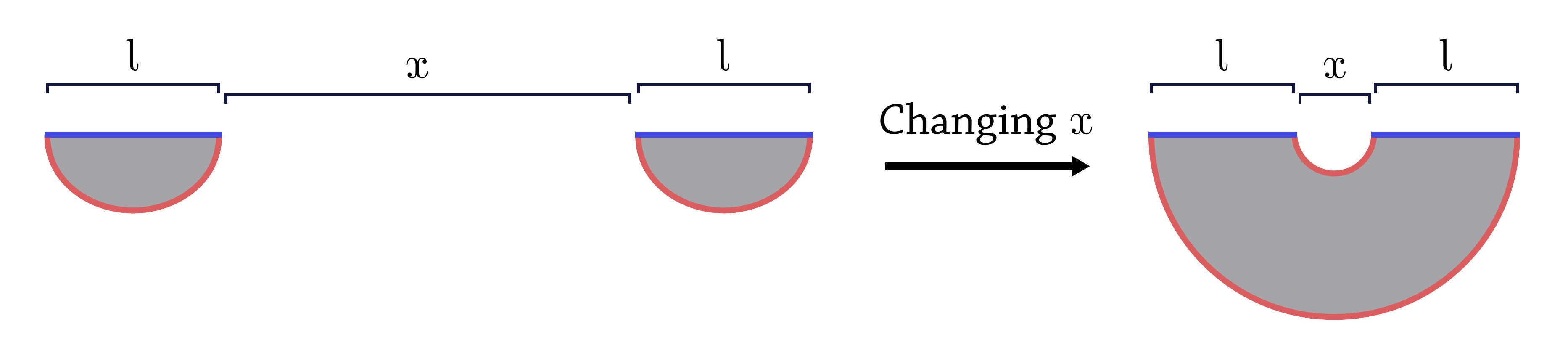}	
	\caption{Changing the separation $x$ between the two strips while keeping the length $l$ fixed (or vice a versa), there is a transition between minimal surfaces marked in red. We also marked in gray the volume inside the minimal surfaces. \label{confutK}}
\end{figure}


\begin{figure}[!h]
	\centering
	\begin{minipage}{0.48\textwidth}
		\centering		
		\includegraphics[width=65mm]{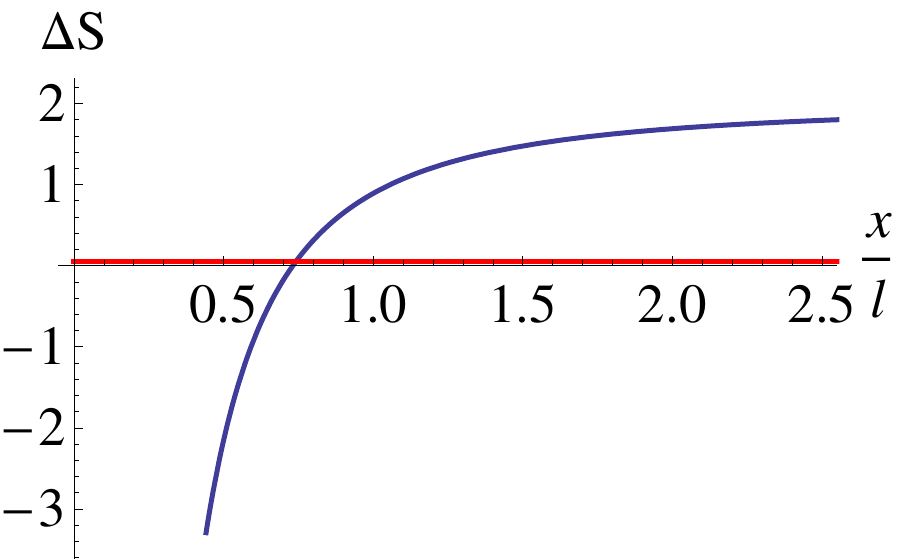}
	\end{minipage}
	\begin{minipage}{0.48\textwidth}
		\centering		
		\includegraphics[width= 65mm]{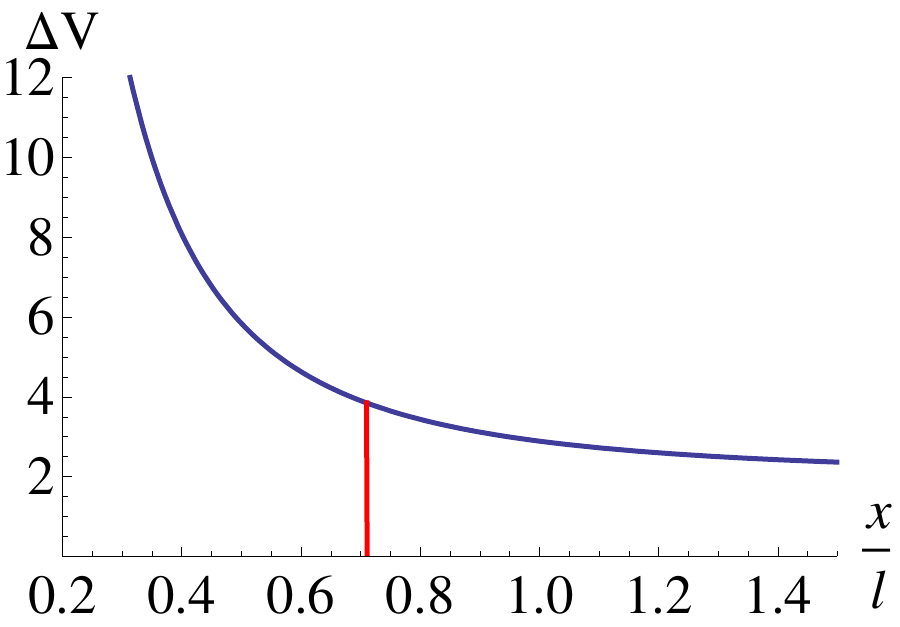}
	\end{minipage}
		\caption{ 2 strips of constant length $l$ for an $AdS_{5}$ geometry.  \textbf{Left:}  $\Delta S=S_{\textrm{con}}-S_{\textrm{disconn}}$ as a function of the  separation distance $x$ (in units of $l$) for a fixed strip length $l$. The red constant line is the disconnected surface area. When $\D S=0$, the EE has a transition, and this occurs at $\frac{x}{l}\approx 0.71$.  \textbf{Right:} Plot of $\Delta V=V_{\textrm{con}}-V_{\textrm{disconn}}$ as a function of the  separation distance $x$ (in units of $l$) for a fixed strip length $l$. From the right plot it is clear that at the transition (marked by the red line) $\Delta V > 0$, hence there will be a jump in the volume. After the transition to the disconnected surface, the area and volume do not depend on $x$. \label{twostripsjump22}}
\end{figure}


When calculating the entanglement entropy of a subregion composed of two disjoint regions, there is a transition between a connected and a disconnected surface. More precisely, as the distance between the subregions is varied, a transition occurs when the areas of the bulk minimal surfaces exchange dominance, see Fig~\ref{confutK}.

The EE of two parallel strips can now be easily calculated. The disconnected surface is simply the addition of two strips of length $l$, and the connected surface is composed of two strips of length $x$ and $2l+x$, see Fig~\ref{confutK}.

When calculating the difference of the volume between two types of surfaces in Fig~\ref{confutK}, the divergent terms cancel. So for clarity we write just the finite terms. The volume of the connected surface is obtained by subtracting the volume of a strip of length $x$ from the volume for a strip of length $(2l+x)$:
\bea
V_{\textrm{con}}= -\frac{c_0}{(x+2l)^{d-1}} +\frac{c_0}{x^{d-1}}
\eea
where $c_0 $ can be read from Eq.~\ref{eq:kjbbbb}.
The volume of the disconnected surface is simply the volume of two strips with length $l$:

\bea
V_{\textrm{discon}}= -\frac{2c_0}{l^{d-1}}
\eea

Subtracting the two volumes, we get:
\bea
\label{eq:ponfll}
\D V =V_{\textrm{con}} - V_{\textrm{discon}}= 
c_0\Big[ -\frac{1}{(2l+x)^{d-2}} + \frac{1}{x^{d-2}} + \frac{ 2}{l^{d-2}}\Big] >0
\eea
Note that this is always positive. Positivity can also be seen from the fact that the disconnected surface is contained inside the connected surface.
In Fig.~\ref{twostripsjump22} we plot $\D S$ and $\D V$ as a function of $\frac{x}{l}$, with fixed $l$ for $AdS_5$ geometry. The right plot shows the finite jump of the volume at the transition.

At the transition point the connected and disconnected surfaces have equal area:
\bea
S_{\textrm{con}} - S_{\textrm{discon}} \Big|_{\textrm{transition}} \propto  - \frac{1}{(2l+x)^{d-2}} - \frac{1}{x^{d-2}} + \frac{ 2}{l^{d-2}} =0
\eea

Plugging this in eq. \ref{eq:ponfll} gives:
\bea
\D V\Big|_{\textrm{transition}}=  V_{\textrm{con}} - V_{\textrm{discon}}\Big|_{\textrm{transition}} = 
\frac{2c_0}{x^{d-2}} >0
\eea

Therefore at the transition point there is a finite jump in the volume. 

Repeating the analysis for $m$ strips gives:
\bea
\D V\Big|_{\textrm{transition}}  = 
(m-1)\frac{2c_0}{x^{d-2}} 
\eea

We can also do the same analysis for a $D_p$-brane geometry with $m$-strips (instead of AdS). This gives (see Eq.~\ref{eq:pkkkde1}):
\bea
\D V\Big|_{\textrm{transition}}  =  
(m-1)\frac{2c_0}{x^{\frac{p+5}{2(5-p)}}} 
\eea


\subsection{Two strips in Global $\textrm{AdS}_3$}

\begin{figure}[!h]
	\centering
	\begin{minipage}{0.48\textwidth}
		\centering		
		\includegraphics[width=52mm]{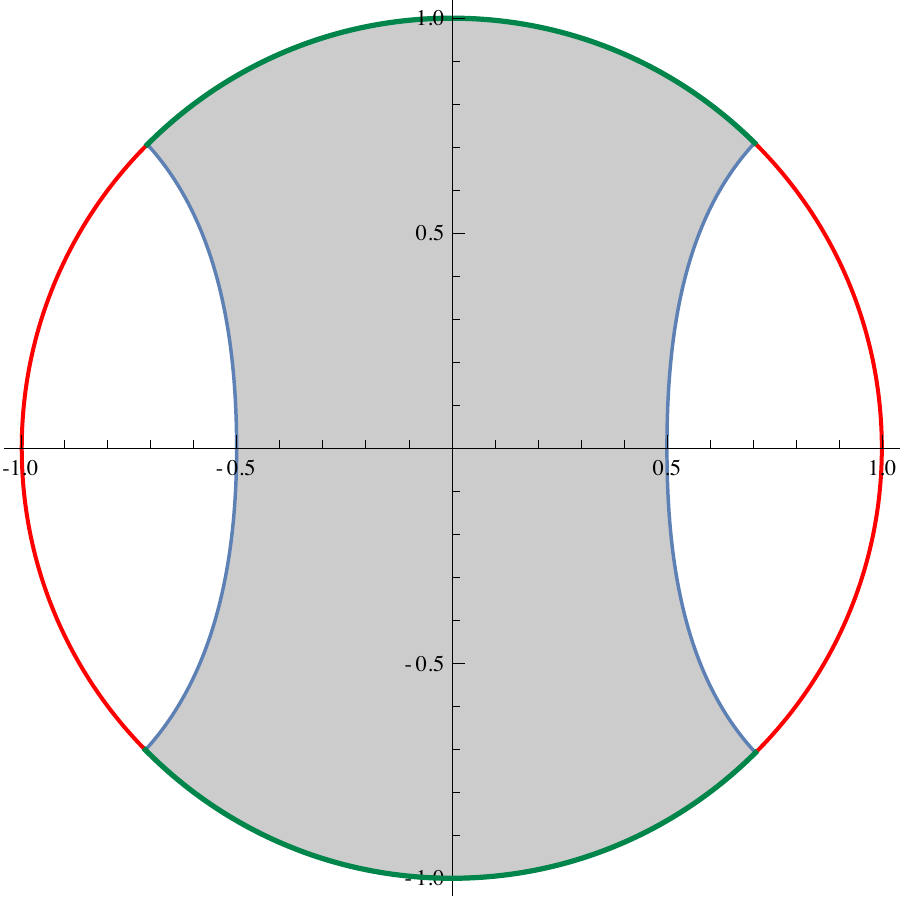}
	\end{minipage}
	\begin{minipage}{0.48\textwidth}
		\centering
		\includegraphics[width= 52mm]{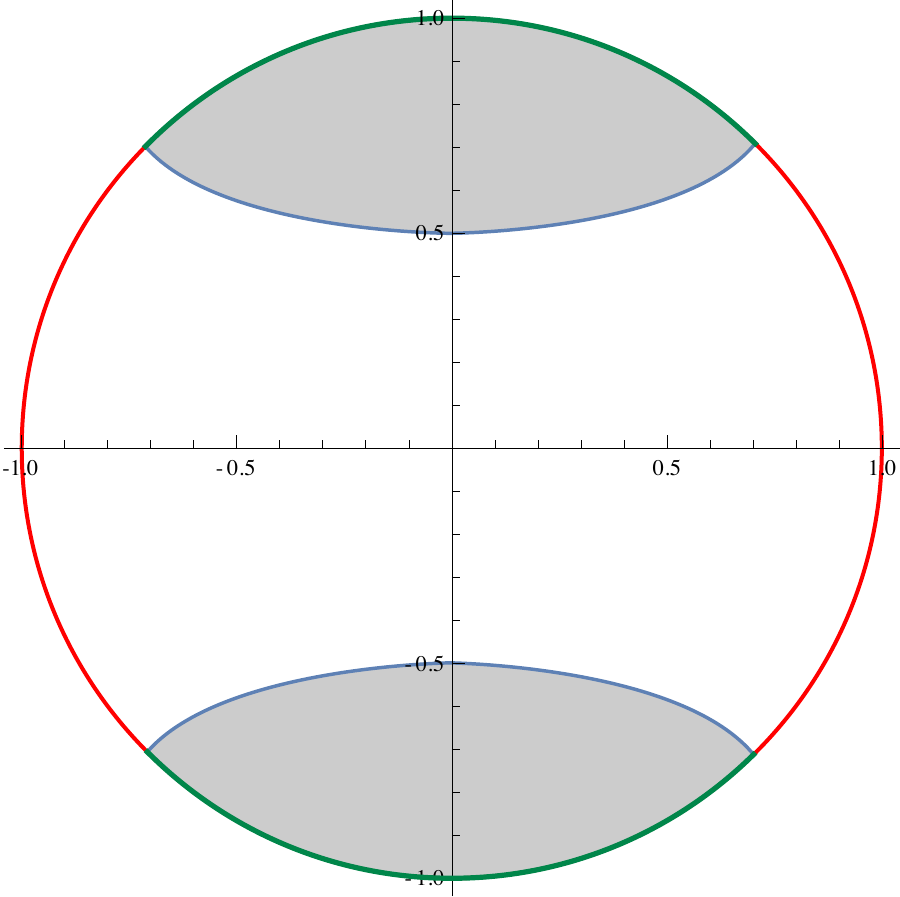}
	\end{minipage}
	\caption{Two strips in global $\textrm{AdS}_3$ at the transition point. The boundary entangling region $A$ is marked in green, the minimal surface in blue and the volume inside the bulk surface in gray. The Left and Right plots show the connected and disconnected minimal surfaces respectively. The difference in volumes between the right and left plots is: $V_{\textrm{conn}}-V_{\textrm{disconn}} = 2\pi$. Note that the disconnected surface is contained inside the connected surface. \label{twostripsads3}}
\end{figure} 

As another example, consider global $AdS_3$ ($d=2$) and a strip entangling region with an opening angle at the boundary $\theta_\infty$. The metric is given by:
\begin{equation}
ds^2=-f(r)dt^2+{dr^2\over f(r)}+r^2d\theta^2, \quad f(r)={r^2\over L^2_{AdS}}+1
\end{equation}
Setting $L_{AdS}=1$ and solving the EOM the geodesic is given by:
\begin{equation}
	r_0\left(\theta\right)=\left({{\sin^2\left(\theta_\infty\right)-\sin^2\left(\theta\right)}\over \cos^2\left(\theta_\infty\right)}\right)^{-{1\over 2}}
\end{equation}

The volume is:
\bea
\label{eq:pokddd}
V(\theta_\infty) =2 \int_0^{\theta_\infty}d\theta \int_{r_0(\theta)}^{r_\infty} dr \frac{r}{(1+r^2)^{\frac{1}{2}}} = 2\theta_{\infty}r_\infty -\pi
\eea 
where $r_\infty= r(\theta_\infty)$ is the UV cutoff. The finite term is a constant equal to $-\pi$.

Now consider 2 strips of opening angle $\theta_\infty$ each.
The ``disconnected" and a ``connected" minimal surfaces are shown in Fig.~\ref{twostripsads3} at their transition point. Their volumes are:
\bea
V_{\textrm{disconn}}(\theta_\infty) = 4\theta_{\infty}r_\infty -2\pi \ \ \ \ \ \ \ \ \ \ \ , \ \ \ \ \ \ \ \ \ \ \ \ V_{\textrm{conn}}(\theta_\infty) = 4\theta_{\infty}r_\infty 
\eea 


Therefore the difference of the volumes is a constant:
\bea
\D V = V_{\textrm{conn}} -V_{\textrm{disconn}}  = 2\pi
\eea 

and the jump in the volume at the transition point is thus:
\bea
\D V\Big|_{\textrm{transition}} = 2\pi
\eea 

\subsection{Strip in a Confining theory dual to $\textrm{AdS}_5\times S^5$ compactified on a circle}
	
\begin{figure}[!h]
	\centering
	\begin{minipage}{0.48\textwidth}
		\centering
		\includegraphics[width= 75mm]{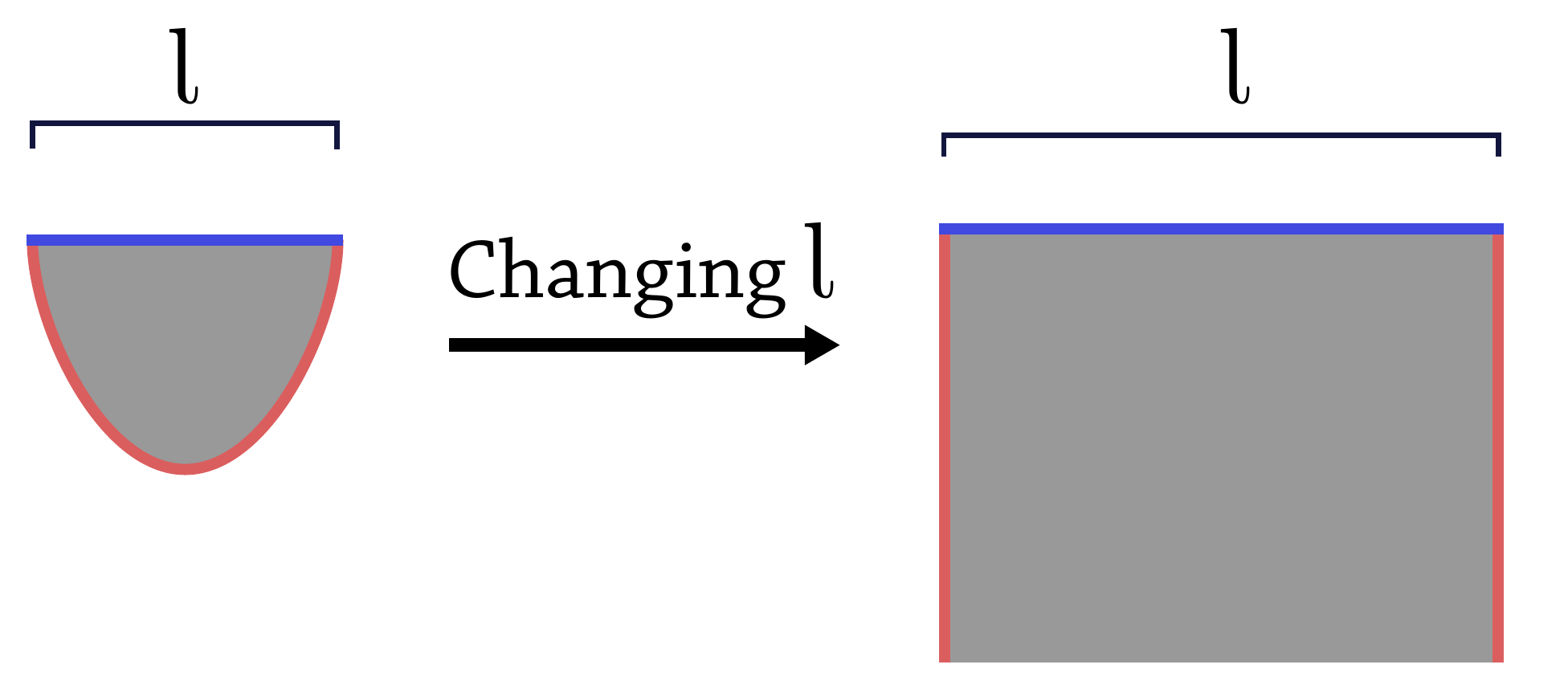}
	\end{minipage}
	\caption{The transition between connected and disconnected minimal surface for a confining background. In red we mark the codim-two minimal surface, in blue the boundary subregion $A$, and in gray the volume inside the minimal surface. \label{confiningonestrip}}
\end{figure} 
\begin{figure}[!h]
	\centering
	\begin{minipage}{0.48\textwidth}
		\centering		
		\includegraphics[width=75mm]{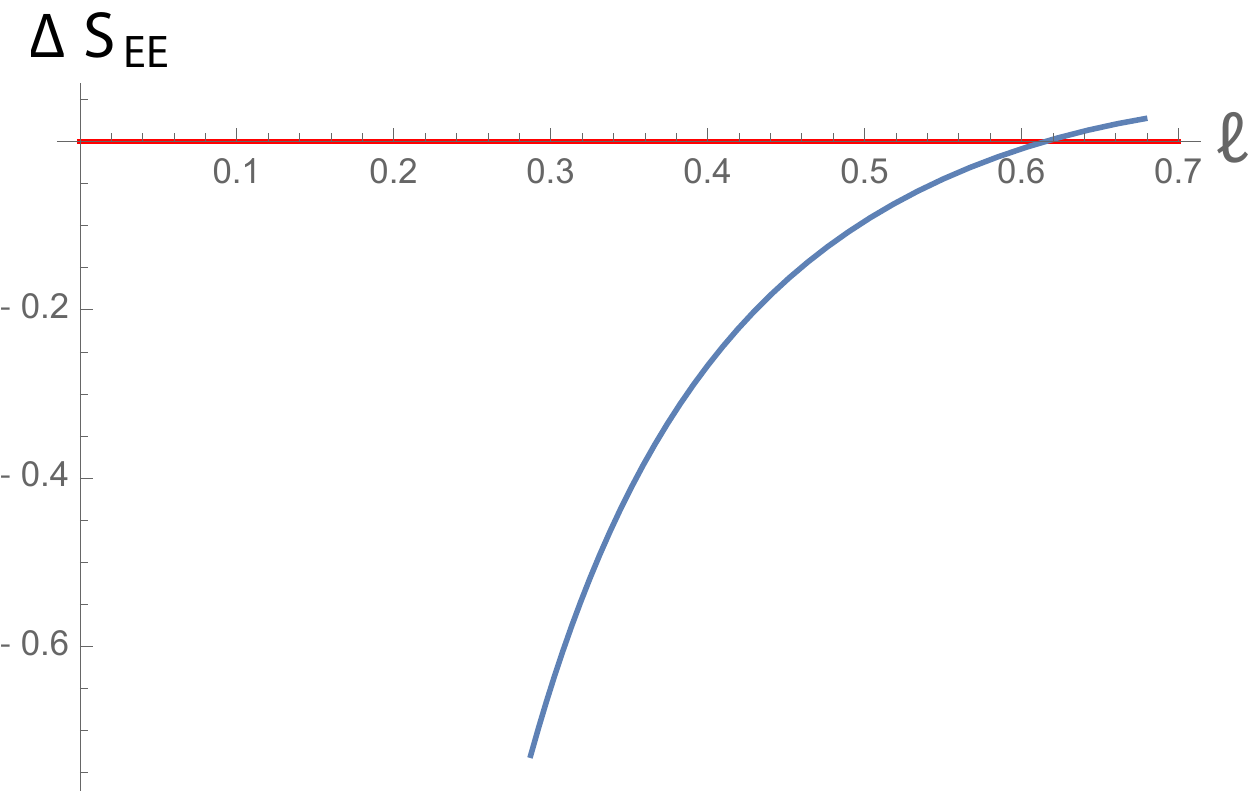}
	\end{minipage}
	\begin{minipage}{0.48\textwidth}
		\centering		
		\includegraphics[width= 75mm]{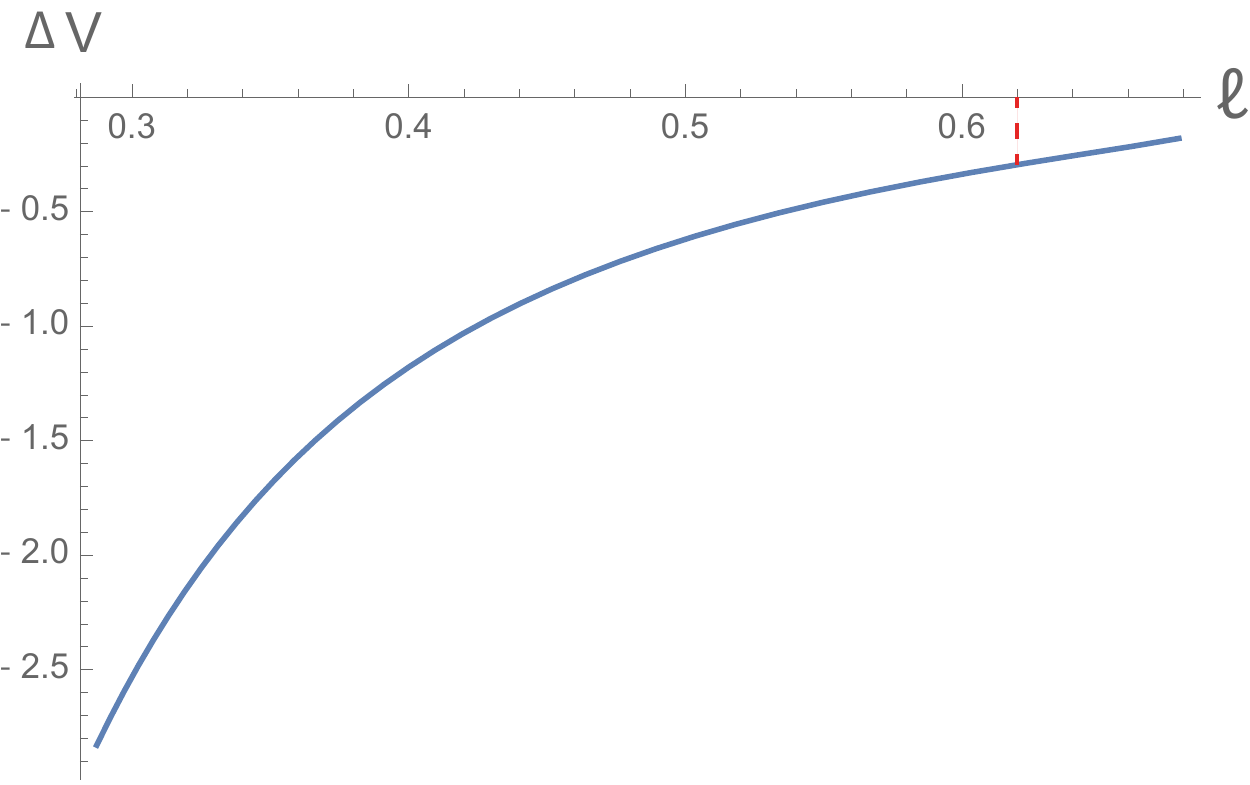}
	\end{minipage}
	\caption{$AdS_5$ compactified on a circle. \textbf{Left:}  $\Delta S_{EE} =S_{\textrm{conn}}-S_{\textrm{disconn}}$ as a function of $l$ (we plot here only the physical branch). The blue and red curves are the areas of the ``connected" and ``disconnected" surfaces respectively. The transition occurs at the point where the curves meet. \textbf{Right:} Plot of $\Delta V=V_{\textrm{conn}}-V_{\textrm{disconn}}$ as a function of $l$. The transition point can be read from the left plot. The transition point is marked by the dashed red line, and it can be seen that there is a jump with $\Delta V < 0$ at the transition point. \label{D9}}
\end{figure}

Another well known example of a holographic entanglement entropy transition occurs for bulk geometries which are dual to confining theories \cite{Witten:1998zw}. These transitions were first studied in \cite{Klebanov:2007ws,Nishioka:2006gr} (see also \cite{Ben-Ami:2014gsa,Kol:2014nqa}), and were interpreted as a probe of the confinement-deconfinement transition. The bulk minimal surface jumps from a ``connected" to a ``disconnected" surface as the the size of the region $A$ is changed, see Fig.~\ref{confiningonestrip}. The entanglement entropy is continuous at the transition, but its derivative is not, as is seen in Fig.~\ref{D9}-Left.

The volume inside the minimal surface has a discontinuous jump at the transition point, as we will now show. We study the case of $\textrm{AdS}_5$ compactified on a spatial circle, with a strip entangling surface. However, we expect a discontinuity to occur generically in any holographic confining theory and any shaped entangling region. Similar to the case of entanglement entropy transitions, we believe that this first order phase transition of the volume is a large $N$ effect, which will be smoothed out after computing $\frac{1}{N}$ corrections \cite{Faulkner:2013ana}. 
 
Consider the metric of $AdS_5$ compactified on a circle in Poincar\'e coordinates:
\begin{equation}
ds^2={\left(L_{AdS}\over z\right)^2}\left[{dz^2\over 1-\left(z\over z_0\right)^4}+dx^\mu dx_\mu\right]+\left(L_{AdS}\over z\right)^2\left(1-\left(z\over z_0\right)^4\right)\left(dx^3\right)^2
\end{equation}
where $L_{AdS}$ is the radius of AdS, and $z_0$ is the point where the contractible cycle shrinks to zero. In this geometry we consider a strip of length $l$.  In  Fig.~\ref{D9} we plot the area difference $\D S_{EE}$ and volume difference $\D V$. From the left plot we see that the transition occurs at $l \approx 0.62$, and from the right plot it can be seen that there is a jump in the volume with $\Delta V < 0$ at the transition point. 
Note that the volume will keep changing (growing linearly with the size $l$ of the region $A$) after the transition to the disconnected surface, unlike the area.

\subsection{Global AdS black holes}	

In this section we consider the global AdS black hole, and a cap entangling surface. We follow the notation of \cite{Hubeny:2013gta}, which considered these ``Entanglement plateaux" transitions for the entanglement entropy. The minimal surface is constrained to satisfy the homology constraint. Additionally, there is a bulk region in which the minimal surface never enters, see Fig. \ref{ads3bhgeod}. As a result there will be a discontinuous volume jump at the transition point, which we now will compute.

Consider the metric for a black hole in global AdS:
\begin{equation}
ds^2=-f(r)dt^2+{dr^2\over f(r)}+r^2\left(d\theta^2+\sin^2\theta d\Omega^2_{d-2}\right) \ \ \ \ \ \ ,\quad \ \ \ \ \ \ \   f(r)=r^2+1-{{r_+^{d-2}\left(r^2_++1\right) \over r^{d-2}}}
\end{equation}
where $r_+$ is the size of the black hole horizon, and we set $L_{AdS}=1$. The relation between the temperature and the size of the horizon is given by $r_+= \frac{2\pi}{d}[RT+ \sqrt{(RT)^2-\frac{d(d-2)}{4\pi^2}}]$ where $R$ is the radius of the boundary sphere.

\subsubsection{Example: BTZ Black hole}

\begin{figure}[!h]
	\centering	
	\begin{minipage}{0.48\textwidth}
		\centering		
		\includegraphics[width=50mm]{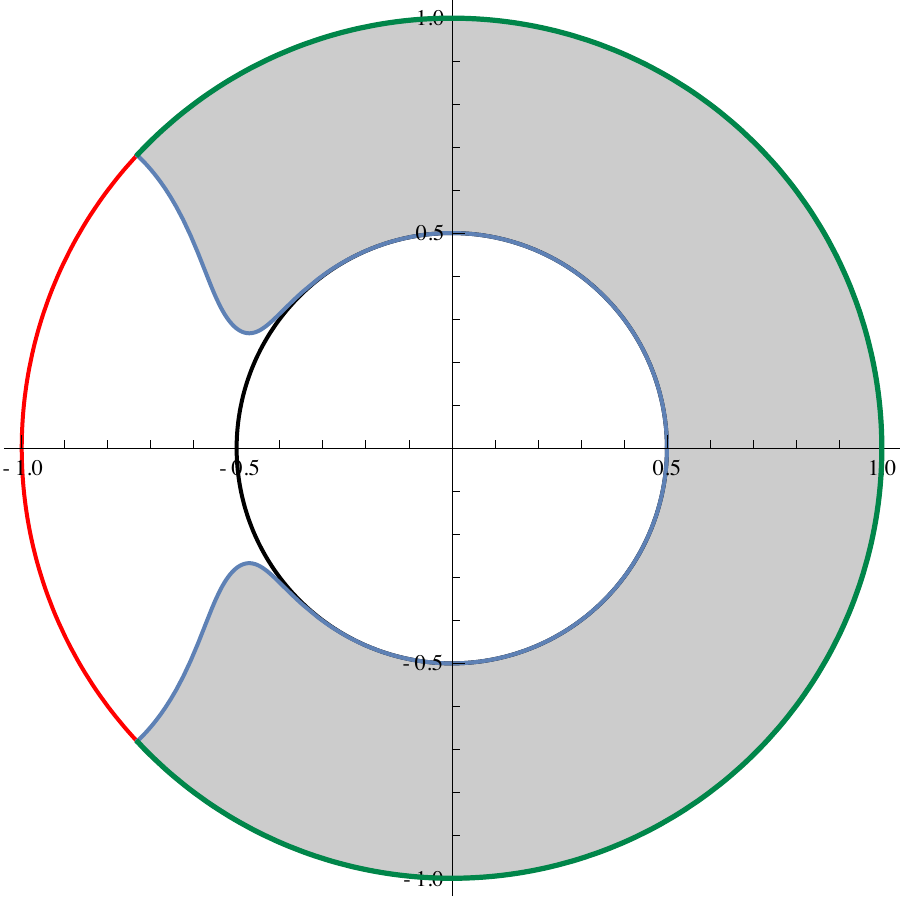}
	\end{minipage}
	\begin{minipage}{0.48\textwidth}
		\centering		
		\includegraphics[width= 50mm]{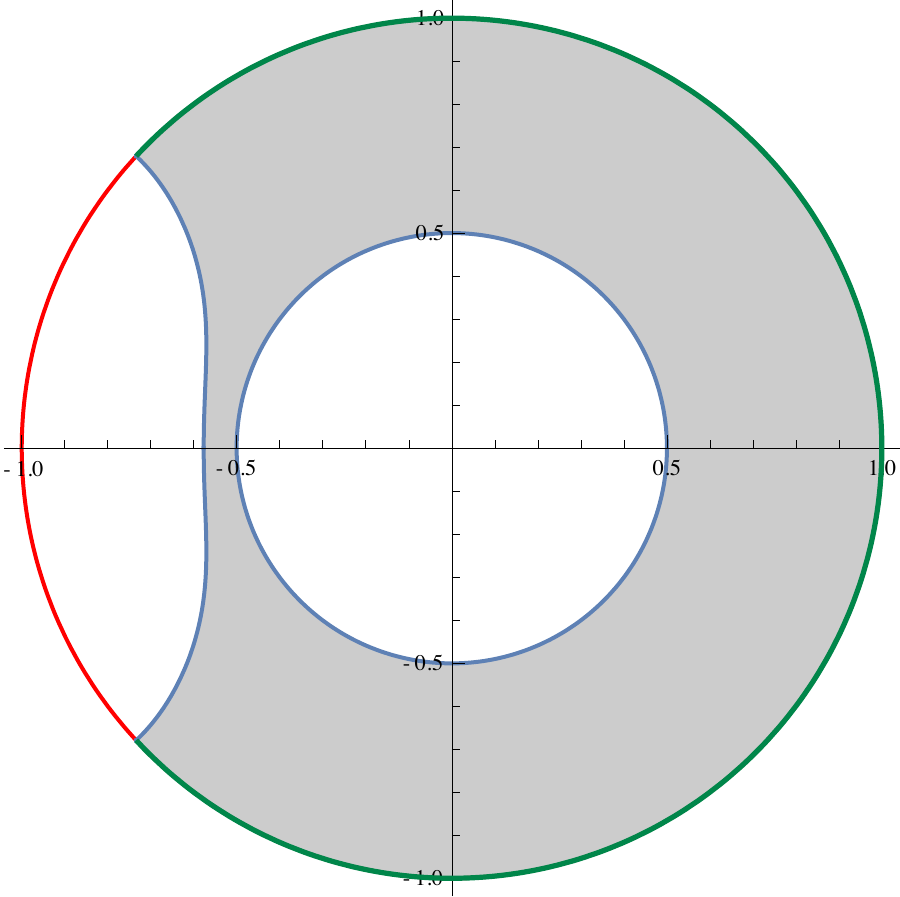}
	\end{minipage}
	\caption{We plot the two types of minimal surfaces for the $AdS_5$ black hole with horizon size $r_+=1$, and a cap entangling region. The surfaces are shown exactly at the transition point (when their areas are equal). In red we mark the boundary, blue is the minimal surface, green is the boundary entangling region, and gray is the volume inside the minimal surface.  \textbf{Left:} The connected minimal surface. \textbf{Right:} The disconnected minimal surface. The homology constraint implies that the horizon is part of the minimal surface.  \label{ads3bhgeod}}
\end{figure}

For the BTZ black hole ($d=2$) we have an analytic solution for the geodesic (see \cite{Hubeny:2013gta}):
\begin{equation}
r\left(\theta\right)=r_+\left(1-{\cosh^2\left(r_+\theta\right)\over \cosh^2\left(r_+\theta_\infty\right)}\right)^{-{1\over 2}}
\end{equation}
where $\theta_\infty$ is the strip opening angle on the boundary.
The minimal surfaces are depicted in Fig.~\ref{ads3bhgeod} at the transition point\footnote{In the figure the horizon size is taken as $r_+=1$.}. The volume for the strip is identical to that of pure AdS, Eq.~\ref{eq:pokddd}:
\bea
V(\theta_\infty) = 2\theta_{\infty}r_\infty -\pi
\eea 

Note that this result is independent of the temperature of the black hole.
Then the difference between the volumes of the connected and disconnected volumes
\bea
\D V = V_{\textrm{conn}}-V_{\textrm{disconn}} = -2\pi
\eea

\begin{figure}[!h]
	\centering	
	\begin{minipage}{0.48\textwidth}
		\centering		
		\includegraphics[width=75mm]{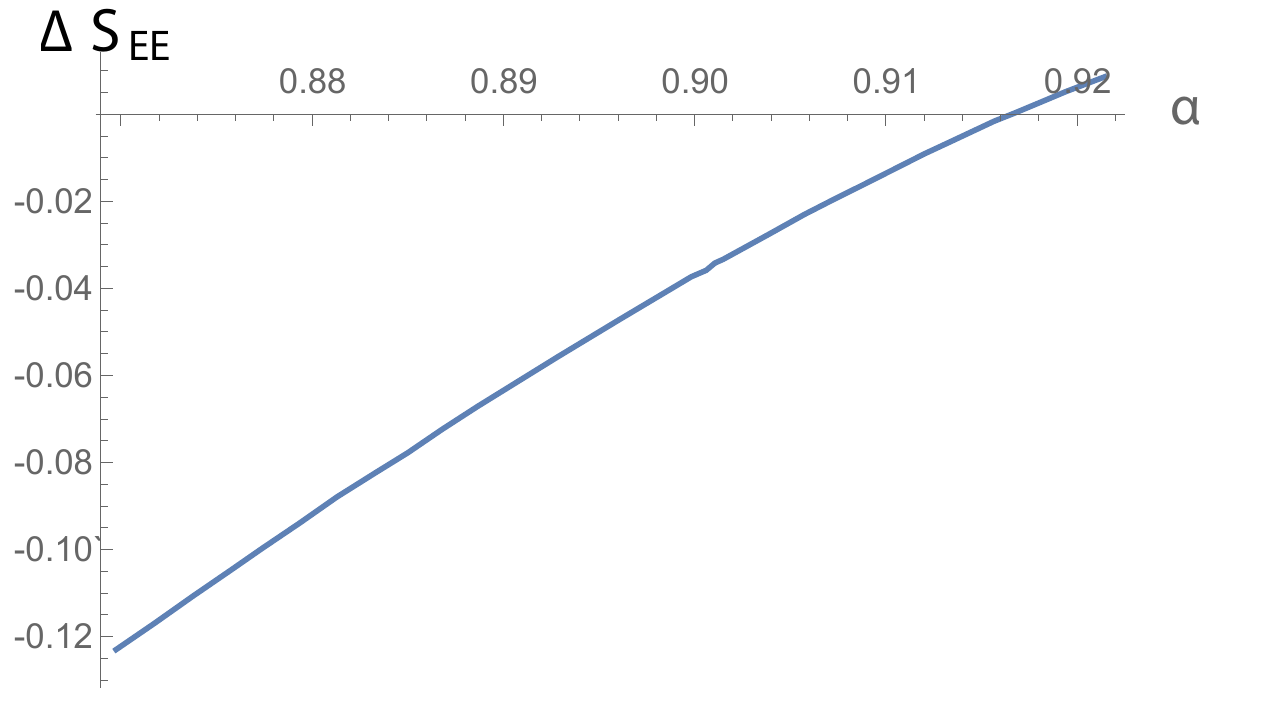}
	\end{minipage}
	\begin{minipage}{0.48\textwidth}
		\centering		
		\includegraphics[width= 75mm]{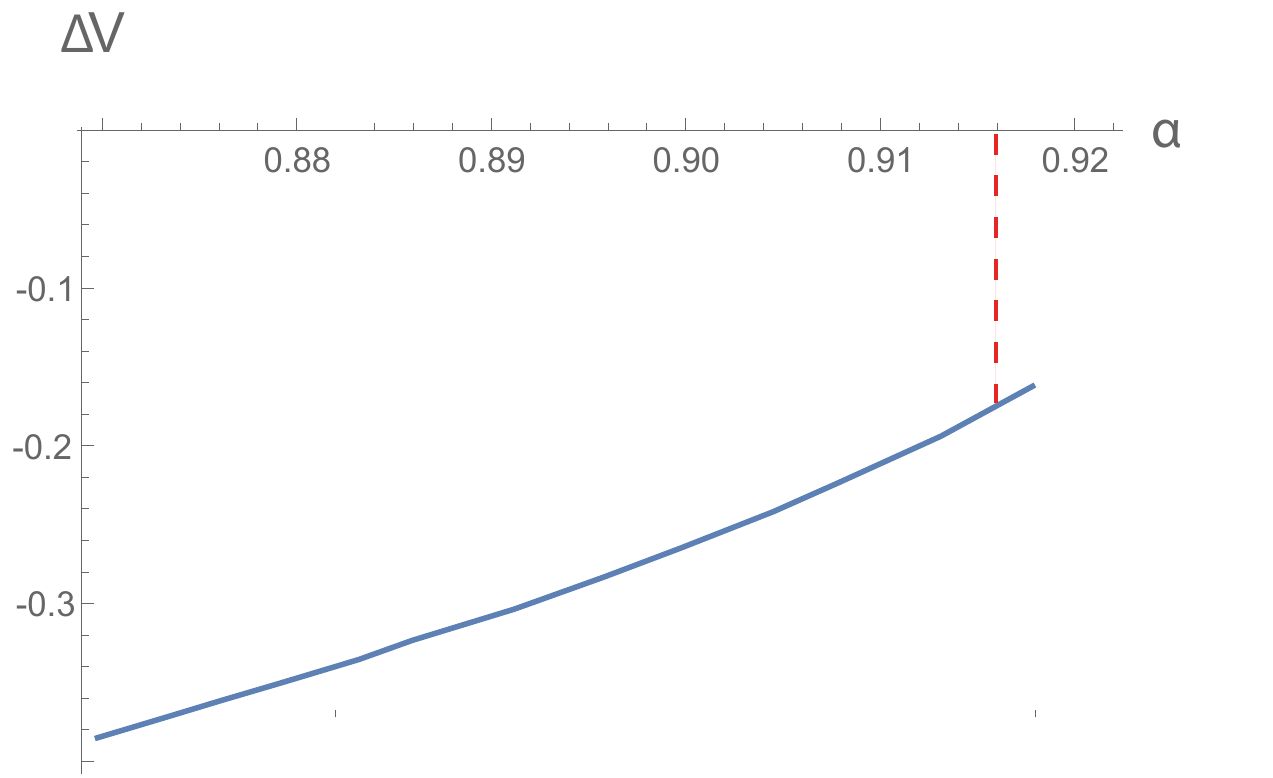}
	\end{minipage}
	\caption{\textbf{Left:} Plot of $\D S_{EE} = S_{\textrm{conn}}-S_{\textrm{discon}}$ as a function of $\a$ for the case of an $AdS_5$ black hole with horizon size $r_+=1$. The transition occurs at $\a \approx 0.915$. \textbf{Right:} Plot of $\D V =V_{\textrm{conn}}-V_{\textrm{disconn}}$ as a function of $\a$. The transition point is marked by the red dashed line, and we see that $\Delta V\big|_{trans.} \approx - 0.18$ in our units. $\label{1g}$}
\end{figure}

\subsubsection{Example: Global $AdS_5$ Black hole}

We numerically calculate the minimal surfaces and the volume inside them (see \cite{Hubeny:2013gta} for more details on the numerical calculation). 
Fig.~\ref{1g}-Left shows the EE difference $\D S_{EE} = S_{\textrm{conn}}-S_{\textrm{disconn}}$ as a function of $\a$, where $\a\equiv\frac{\textrm{vol}(\mm{A})}{\textrm{vol}(S^{d-1})}$ (the size of the entangling surface which is a cap centered at the $\theta=0$ with radius $\theta_\infty$). One can see a transition at $\a \approx 0.915$ when $\D S_{EE} =0$.
Fig.~\ref{1g}-Right shows the volume difference $\D V = V_{\textrm{conn}}-V_{\textrm{disconn}}$ as a function of $\a$. One can see that the volume of the disconnected surface is always larger than the connected as expected. In particular, at the transition point ($\a \approx 0.915$) the volume jumps discontinuously: $\Delta V\big|_{transition} \approx - 0.18< 0$ in our units..


\section{Discussion}	
\label{sec:discuss}

In this paper we studied the codim-one volume inside the Ryu-Takayanagi surface, first considered in \cite{Alishahiha:2015rta}. We performed explicit computations for various examples in time-independent geometries. We found that for an AdS black hole, the volume is a non-monotonic function of the strip's length at a fixed finite temperature, even though the EE is monotonic. We also showed that the volume exhibits a discontinuous jump in geometries for which two bulk minimal surfaces exchange dominance. It will be important to understand better the QFT quantity dual to the subregion volume, and the significance of its universal terms in order to compare to the bulk results. It will be also interesting to understand the dual interpretation of the discontinuous jump in the volume assuming that this is indeed the dual of quantum complexity of a subregion. As we mentioned before these jumps imply that the volume of the entanglement wedge exhibits a discontinuous jumps under small changes of the reduced density matrix.


It would be interesting to compute the subregion volume for time dependent backgrounds (e.g quenches). The time dependence was very important in the study of holographic complexity of Susskind et al. for the case in which the region $A$ is the whole boundary, and the bulk geometry is a double sided black hole. In that case, the linear late time behavior matches the expectation from complexity.

It would also be interesting to understand $\frac{1}{N}$ corrections to the subregion volume. For the entanglement entropy this is given by the bulk EE, where the entangling region is precisely the volume inside the RT surface \cite{Faulkner:2013ana}. It would be interesting to understand if there is a relation to bulk relative entropy \cite{Jafferis:2015del,Jafferis:2014lza,Blanco:2013joa}. Another direction would be to understand the behavior of the volume for higher derivative gravity in the bulk.

The suggestion of \cite{Freedman:2016zud} is a reformulation of the Ryu-Takayanagi formula in terms of bit threads, which loosely speaking  probe the volume inside the RT surface. It would be interesting to understand our results in terms of the bit thread formulation.

In \cite{Lashkari:2015hha} it was shown that the quantum fisher information metric (the 2nd correction of the relative entropy) for a sphere and a CFT is dual to the canonical energy integrated inside the RT surface. Instead, one can alternatively integrate other bulk fields inside the volume, and try to understand the dual information theoretic quantities.

One can study further properties of the subregion volume by performing relevant perturbations or shape perturbations. From the analysis of \cite{Carmi:2015dla}, it can be immediately shown that the volume corresponding to entangling surfaces with a rotational or translational symmetry is an extremum with respect to shape perturbations of the entangling surface.  

For some entangling surfaces $\S$, when there are no transitions between bulk extremal surfaces, the volume inside the RT surface can be foliated with  bulk extremal surfaces attached to concentric entangling surfaces with different scales. Perhaps this result can relate the subregion volume (complexity) to an integral over entanglement entropies.\\

\textbf{Acknowledgments}

We thank Rob Myers for many helpful discussions. We also thank Mohsen Alishahiha, Michal Heller, and Pratik Rath.  The work of OB and DC is partially supported by the Israel
Science Foundation (grant 1989/14 ), the US-Israel bi-national fund (BSF) grant 2012383
and the German Israel bi-national fund GIF grant number I-244-303.7-2013. DC is grateful for the support from the ``visiting graduate fellows" program at the Perimeter Institute . Research at Perimeter Institute is supported by the Government of Canada through the Department of Innovation, Science and Economic Development and by the Province of Ontario through the Ministry
of Research \& Innovation.


\appendix

\section{Leading temperature Correction for a Sphere}
\label{sec:sphere}

We will show that the 1st order temperature correction to the volume vanishes for a sphere entangling surface of radius $l$ (we thus verify the calculation of \cite{Alishahiha:2015rta}). 
Consider again the general asymptotically AdS metric:
\bea
ds^2 =\frac{1}{z^2}\Big[ f_0(z)dt^2 + f_1(z) dx_\m^2 + f_2(z) dz^2  \Big] 
\eea  
Where for a black hole geometry at small temperature, $f_1(z)=1$ and  $f_2(z)= \frac{1}{1- k_0T^d z^d} \approx 1+k_0T^dz^d$.

The area of a bulk surface is:
\bea
\textrm{Area}= \int_{\d}^{z_*} dz \frac{r(z)^{d-2}\sqrt{f_2(z)+r'^2}}{z^{d-1}} 
\eea 
where $r$ is a radial coordinate.
The equation of motion is:
\bea
(d-2)\frac{r^{d-3}}{z^{d-1}}\sqrt{f_2(z)+r'^2} - \frac{d}{dz}  \Big(\frac{r^{d-2}r'}{z^{d-1}\sqrt{f_2(z)+r'^2}} \Big)=0
\eea 

To 1st order in $T^d$, the solution is:
\bea
\label{eq:sdjlkklh}
r(z)= \sqrt{z_*^2-z^2} +k_0T^d \frac{2z_*^{d+2}-z^d(z_*^2+z^2)}{2(d+1)\sqrt{z_*^2-z^2}}
\eea 

The volume inside the bulk minimal surface is:
\bea
V= \int_{\d}^{z_*}dz \frac{ \sqrt{f_2(z)}r^{d-1}(z)}{z^{d}} 
\eea

Plugging $r(z)$ from Eq.~\ref{eq:sdjlkklh} gives:
\bea
V(l)\propto \int_{\d}^{l-\frac{k_0T^d l^{d+1}}{(d+1)} }dz \frac{(l^2-z^2)^{(d-1)/2}}{z^{d}}  \Big[1+\frac{k_0T^d}{2}z^d\Big]\Big[1 -k_0T^d \frac{(d-1)z^d(l^2+z^2)}{2(d+1)(l^2-z^2)} \Big]
\eea 

Expanding at linear order in $T^d$, and changing variables to $t=z/l$ gives:
\bea
V(l)- V_{T=0}(l) \propto  \frac{T^dl^d}{(d+1)}\int_{0}^{1 }dt \frac{(1-t^2)^{\frac{d-3}{2}}}{t^{d}}   \Big[t^d - dt^{d+2} \Big]  =0
\eea 

and the integral is zero. Therefore we showed that the 1st order temperature correction to the volume vanishes for a sphere entangling surface.


\section{The Volume and Action of the Entanglement wedge}
\label{sec:wedge3}


The entanglement wedge \cite{Headrick:2014cta} is a codim-zero bulk region defined as the bulk domain of dependence\footnote{The causal domain of dependence $D[A]$ of a region $A$ is defined as the set of all points $P$ for which all causal curves through $P$ intersect $A$.} of the volume inside the HRT surface. Thus the HRT surface is the rim of the entanglement wedge. It was was argued in \cite{Headrick:2014cta} that the entanglement wedge is the region most suitable to be the bulk dual of the reduced density matrix. 








\subsection{The volume of the entanglement wedge for a sphere and pure AdS}

As a simple example consider Poincar\'e $AdS_{d+1}$ and a sphere entangling surface of radius $l$:
\bea
z^2+r^2 = l^2
\eea 

\begin{figure}[!h]
	\centering
	\includegraphics[width=90mm]{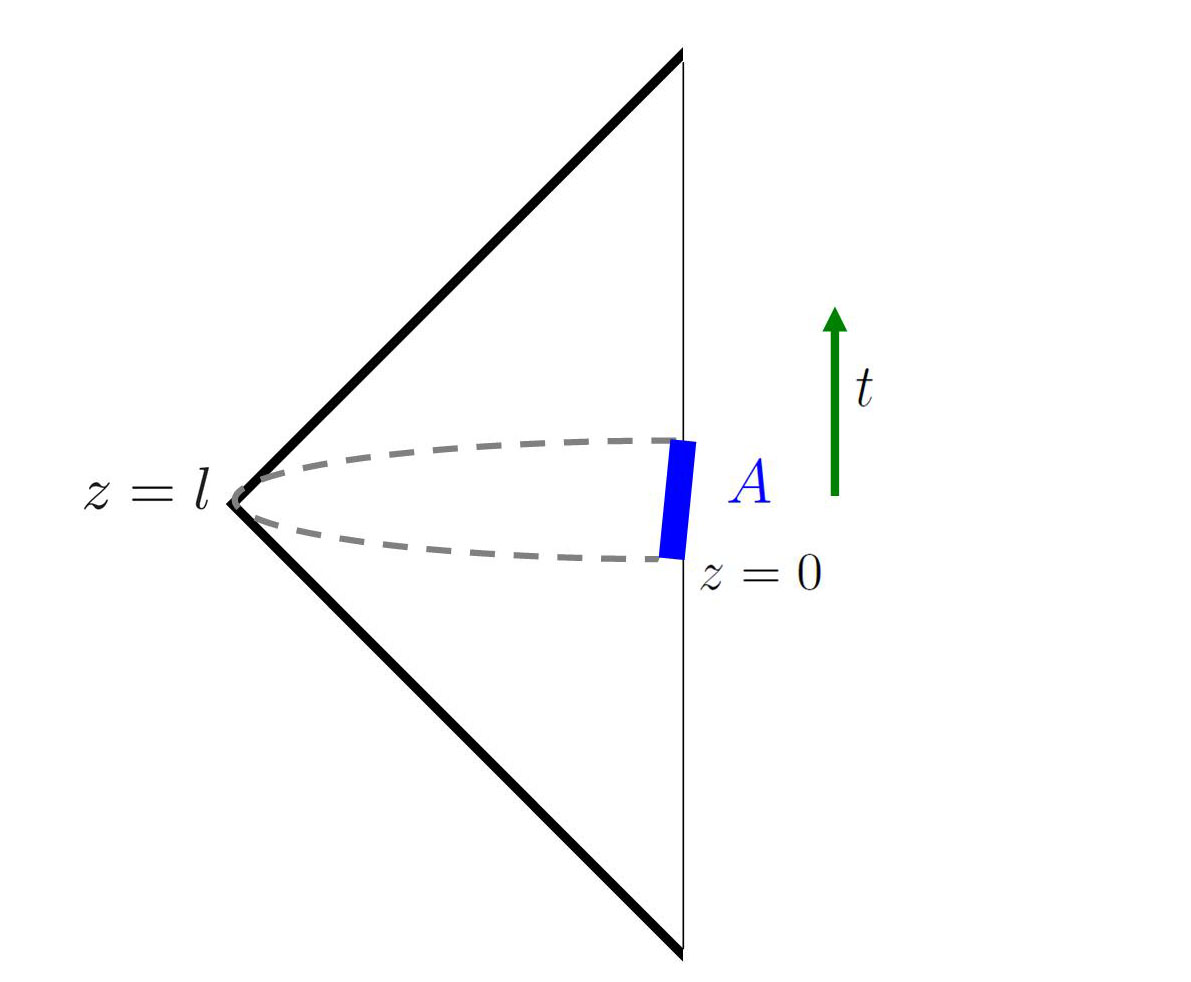}	
	\caption{ For $AdS_3$ and an interval $A$, showing the the entanglement wedge region (the inside of the half-cone). The region $A$ is the blue interval, and the bulk minimal surface is the gray dashed curve. \label{Wedge2}}
\end{figure}

The corresponding entanglement wedge is half of a cone in the bulk in Poincar\'e coordinates, and is shown in Fig.~\ref{Wedge2} for the case of $AdS_3$. The null boundary surface of the entanglement wedge (for $t>0$) is given by the equation:
\bea
\label{eq:pkdddk}
\sqrt{r^2+z^2}=l-t
\eea 
where $t$ is the time,  $r=\sqrt{x_i^2}$ is a boundary radial coordinate, $z$ is the bulk coordinate, and $l$ is the radius of the entangling surface sphere.
Thus, the volume of the entanglement wedge is simply given by twice (because there is an equal contribution from $t<0$) the volume inside the surface of Eq.~\ref{eq:pkdddk}:
\bea
\label{eq:pkdddk2}
V_{wedge} =2 \int_0^{l} dt \int_\d^{l-t} dz\sqrt{g(z)} \int_0^{\sqrt{(l-t)^2-z^2}} dr r^{d-2} \O_{d-2} 
\eea 
where $\O_{d-2}$ is the area of a $(d-2)$-sphere, and $ \sqrt{g(z)}=z^{-(d+1)}$. We can easily expand this integral to derive the leading divergences:
\bea
\label{eq:vwegem}
V_{wedge} =\frac{2\O_{d-2} }{d-1} \int_0^{l} dt \int_\d^{l-t} dz \frac{((l-t)^2-z^2)^{\frac{d-1}{2}}}{z^{d+1}} =
\frac{2\O_{d-2} }{d-1} \Big[ \frac{1}{d^2}\frac{l^d}{\d^d} - \frac{d-1}{2(d-2)^2}\frac{l^{d-2}}{\d^{d-2}} + \ldots \Big]
\eea 

Note that the leading divergence goes like $\frac{1}{\d^d}$, and there are subleading divergences $\frac{1}{\d^{d-2}}+\ldots$ etc. We can also easily compute the universal finite or log terms. 

\subsection{The action in the entanglement wedge}

In the previous section we calculated the volume of the entanglement wedge. In this section we consider a related quantity, the gravitational action computed inside the entanglement wedge. We stick to the simple example of a sphere entangling surface and Poincar\'e AdS.

The gravitational action is composed of a bulk term, a null-boundary term, and a corner term (for more details see \cite{Lehner:2016vdi,divergences,Parattu:2015gga}):
\bea
\mm{A}&=\mm{A}_{bulk} + \mm{A}_{Null}+ \mm{A}_{corner}= 
\frac{1}{16\pi G} \int d^{d+1}x\sqrt{-g}\Big(\mm{R}-\frac{d(d-1)}{L_{AdS}^2}\Big) \\ 
&+\frac{1}{8\pi G} \int d\l \sqrt{-\g} d^{d}x  \k + \frac{1}{8\pi G}\int d^{d-1}x \sqrt{-\tilde g}a 
\eea 

For pure AdS, the bulk term is proportional to the volume that we computed in a previous section:
\bea
\mm{A}_{bulk} =\Big(\mm{R}-\frac{d(d-1)}{L_{AdS}^2}\Big)V_{wedge}
\eea 
where $V_{wedge}$ was computed in Eq.~\ref{eq:vwegem}.

We can choose an affine parameter, and then $\k$ is zero and the null boundary terms vanish, see \cite{Lehner:2016vdi,divergences,Parattu:2015gga}:
\bea
\mm{A}_{Null} =0 
\eea 

The codim-two corner term of the action arises from the dashed gray curve in Fig.~\ref{Wedge2}. The action contribution from this corner is (see \cite{Lehner:2016vdi,divergences}) :
\bea
\mm{A}_{corner} =
\log \Big(\frac{\sqrt{u_3 u_4}}{L^2} \d \Big) \Big[  \frac{l^{d-2}}{\d^{d-2}} +  \frac{l^{d-4}}{\d^{d-4}}  + \ldots \Big] +\Big[  \frac{l^{d-2}}{\d^{d-2}} +  \frac{l^{d-4}}{\d^{d-4}}  + \ldots \Big]
\eea 
where $u_{3,4}$ are the normalization vectors of the null boundaries of the entanglement wedge. Above, there are only even or odd powers, but not both.
Summing the bulk and corner terms, we obtain the action in the entanglement wedge.





\bibliographystyle{utphys}

\bibliography{finite}

\end{document}